\documentclass{article}
\usepackage{amsmath}
\usepackage{amssymb}
\usepackage{graphicx}
\usepackage{hyperref}
\usepackage{xcolor} 
\usepackage{adjustbox} 
\usepackage{booktabs} 
\usepackage{caption} 
\usepackage{tocloft} 
\usepackage{array}    
\usepackage{siunitx}  
\usepackage{float} 
\usepackage{abstract}
\usepackage[paper=a4paper]{geometry}
\usepackage{pdfpages}
\usepackage{tikz}
\usetikzlibrary{arrows.meta} 
\usepackage{mathrsfs}
\usepackage{authblk}       
\usepackage{hyperref}      
\usepackage{multirow}

\hypersetup{
    colorlinks=true,
    linkcolor=blue,
    filecolor=magenta,      
    urlcolor=cyan,
}

\title{A Review of Graph Neural Network Models in Epidemic Forecasting: EpiGNN and ColaGNN}

\makeatletter

\def\@fnsymbol{\@arabic}
\makeatother

\title{\Large\bfseries Computation for Epidemic Prediction with Graph Neural Network by Model Hybridization}
\date{}

\author[1]{Xiangxin Kong}
\author[2]{Hang Wang}
\author[3]{Yutong Li}
\author[4]{Yanghao Chen}
\author[5,*]{Zudi Lu}

\affil[1]{Department of Biostatistics, City University of Hong Kong; xikong9-c@my.cityu.edu.hk}
\affil[2]{Department of Biostatistics, City University of Hong Kong; hwang858-c@my.cityu.edu.hk}
\affil[3]{Department of Biostatistics, City University of Hong Kong; yli2487-c@my.cityu.edu.hk}
\affil[4]{Department of Biostatistics, City University of Hong Kong; yanghchen5-c@my.cityu.edu.hk}
\affil[5]{Department of Biostatistics, City University of Hong Kong; \textit{Correspondence:}~\texttt{zudilu@cityu.edu.hk}}

\begin{document}

\maketitle

\begin{abstract}
Modelling epidemic events such as COVID-19 cases in both time and space dimensions is an important but challenging task. Building on in-depth review and assessment of two popular graph neural network (GNN)-based regional epidemic forecasting models of \textbf{EpiGNN} and \textbf{ColaGNN}, we propose a novel hybrid graph neural network model, \textbf{EpiHybridGNN}, which integrates the strengths of both EpiGNN and \textbf{ColaGNN}. In the EpiGNN, through its transmission risk encoding module and Region-Aware Graph Learner (RAGL), both multi-scale convolutions and Graph Convolutional Networks (GCNs) are combined, aiming to effectively capture spatio-temporal propagation dynamics between regions and support the integration of external resources to enhance forecasting performance. While, in the ColaGNN,  a cross-location attention mechanism, multi-scale dilated convolutions, and graph message passing are utilized to address the challenges of long-term forecasting through dynamic graph structures and spatio-temporal feature fusion. Both enjoy respective advantages but also share mutual shortcomings.  Our EpiHybridGNN is therefore designed to combine the advantages of both EpiGNN, in its risk encoding and RAGL, and ColaGNN, in its long-term forecasting capabilities and dynamic attention mechanisms.  This helps to form a more comprehensive and robust prediction of spatio-temporal epidemic propagation. The computational architecture, core formulas and their interpretations of our proposed EpiHybridGNN are provided. Multiple numerical real data experiments  validate that our EpiHybridGNN significantly outperforms both EpiGNN and ColaGNN in epidemic forecasting with comprehensive insights and references offered.


\end{abstract}

\section{Introduction}
The global spread of epidemics poses a significant threat to public health. Accurate epidemic forecasting is crucial for effective disease control, optimizing medical resource allocation, and promoting drug development. Traditional models\cite{alabdulrazzaq2021accuracy, kane2014comparison, wang2015dynamic}, due to overly simplistic assumptions, struggle to capture complex non-linear spatio-temporal relationships. In recent years, deep learning, especially Graph Neural Networks (GNNs)\cite{bronstein2017geometric, zhou2018gnnreview}, has emerged as a vital tool for epidemic forecasting due to its powerful ability to process non-Euclidean structured data and capture complex correlations between nodes.

It is against this backdrop that EpiGNN and ColaGNN emerged as innovative GNN models. EpiGNN focuses on integrating transmission risk (including local and global transmission risk) and geographical information, emphasizing explicit modeling of local and global propagation mechanisms. Through its Region-Aware Graph Learner (RAGL) and multi-scale convolutions, it supports external resource integration to enhance forecasting performance. ColaGNN, conversely, emphasizes long-term forecasting through a cross-location attention mechanism and multi-scale dilated convolutions, excelling at capturing complex dynamic dependencies of non-adjacent regions, particularly in long-term predictions.

Despite their individual strengths in specific scenarios, EpiGNN and ColaGNN exhibit significant complementarity. EpiGNN excels in external data integration and handling early and mid-term changes, but its graph construction mechanism limits its long-term forecasting ability. ColaGNN performs outstandingly in long-term forecasting but may lack explicit risk encoding.

This paper aims to provide an in-depth analysis and comparison of EpiGNN and ColaGNN, and building on this, proposes a novel hybrid model named HybridGNN (EpiCola-GNN). HybridGNN stands out with its innovative modular architecture, designed to excel in modeling both short-term and long-term disease propagation patterns. The model begins with a Temporal Feature Extraction module, inspired by ColaGNN, which employs multi-scale 1D convolutions to extract short- and long-term temporal features. This is complemented by a Transmission Risk Encoding module, drawing from EpiGNN, which computes local risk using geographic degree information and global risk via temporal features. A key highlight is the Dynamic Graph Construction module, which fuses ColaGNN’s RNN-based attention mechanism with EpiGNN’s geographic and external relation priors to generate a dynamic adjacency matrix. This dynamic framework supports a Feature Fusion and Graph Message Passing process through multi-layer GCNs, where initial node features are propagated and enriched with residual connections. Finally, the Prediction module concatenates GCN outputs with RNN hidden states, optionally incorporating a residual window to yield the final prediction. By leveraging ColaGNN’s efficient stable long-term forecasting performance with EpiGNN’s short-term dynamic modeling, HybridGNN achieves a balanced approach that excels in capturing rapid outbreaks as well as prolonged epidemic trends. This dual-focus capability, combined with its scalability, positions HybridGNN as a powerful tool for epidemic forecasting.

The remainder of this paper is organized as follows. This paper begins by reviewing related work in Section 2, followed by a detailed description of the EpiGNN and ColaGNN architectures in Section 3. Section 4 presents the experimental analysis of the two models. Based on the comparative advantages and disadvantages discussed in Section 5, we propose HybridGNN in Section 6 to enhance performance across multiple datasets. Then we analyze the strengths of the proposed HybridGNN framework and discuss potential directions for further improvement in Section 7. Finally, conclusions are drawn in Section 8.

\section{Related Work}
In the field of epidemic forecasting, recent studies have increasingly leveraged deep learning models. For instance, research by \cite{venna2018novel, wu2018deep} specifically explored models for direct long-term epidemiological predictions, achieving good results. Wu et al. \cite{wu2018deep} proposed CNNRNN-Res, a model combining CNN and RNN. DEFSI \cite{wang2019defsi} integrated deep neural network methods with causal models. Jung et al. \cite{jung2021self} designed a self-attention-based method for regional influenza forecasting.

GNNs have shown prominent performance in epidemic forecasting due to their ability to capture inter-node correlations in graph structures. For example, STGCN \cite{yu2017spatio} is a deep learning model specifically designed to process data with spatial and temporal dependencies. It combines Graph Convolutional Networks (GCNs) \cite{zhang2019gcnreview, kipf2016semi} with traditional sequence models (such as convolutional sequence learning or recurrent neural networks) to simultaneously capture spatial and temporal features in the data. However, existing GNN methods face two major challenges: accurately capturing potential propagation relationships between regions and effectively utilizing regional propagation risk information. Both EpiGNN and ColaGNN address these challenges effectively through innovative module designs. Through the Region-Aware Graph Learner (RAGL), EpiGNN \cite{xie2022epignn} can dynamically learn and fuse temporal correlations, geographical priors, and external resources to construct an adaptive propagation graph structure. This enables flexible capture of complex and evolving propagation paths. It extracts multi-scale temporal features via Multi-Scale Convolutions, and combines them with Local Transmission Risk Encoding and Global Transmission Risk Encoding to comprehensively encode a region's potential propagation risk. With powerful dynamic graph learning capabilities, EpiGNN is highly effective at capturing rapidly evolving, non-fixed propagation relationships in the short term, but potentially leading to relative deficiencies in modeling long-term, stable temporal dependencies. Through its Loc-Aware Attention, ColaGNN \cite{deng2020cola} dynamically calculates and adjusts edge weights between regions on top of a pre-defined base graph structure using an attention mechanism. This allows for more adaptive information propagation and the measurement of influence between non-adjacent regions. The combination of RNN and Dilated Conv modules makes it highly effective at capturing long-term, complex temporal patterns. Despite introducing dynamic edge weights, ColaGNN's underlying graph topology fundamentally relies on pre-defined connections. For propagation events that change rapidly in the short term or completely break existing topological structures, its adaptability is less robust compared to EpiGNN. Therefore, to construct a more comprehensive and robust spatio-temporal forecasting model that can effectively address both short-term dynamics and long-term trends, it is necessary to build a HybridGNN.

\section{Model Architecture}
\subsection{Problem Definition}
We formulate epidemic forecasting as a graph-based spatiotemporal propagation problem. Let there be $N$ locations. The historical case data is represented as $\mathbf{X} = [\mathbf{x}_1, \dots, \mathbf{x}_t]$, where each $\mathbf{x}_z \in \mathbb{R}^N$ denotes the case counts across all locations at time $z$. The goal is to predict future case counts $\mathbf{x}_{t+h}$, where $h$ is the prediction horizon. We utilize a look-back window of size $T$, i.e., $[\mathbf{x}_{t-T+1}, \dots, \mathbf{x}_t] \in \mathbb{R}^{N \times T}$, as input. For each region $i$, the corresponding time series is denoted as $\mathbf{x}_{i:} = [x_{i,t-T+1}, \dots, x_{i,t}] \in \mathbb{R}^{1 \times T}$. In addition, we incorporate a static geographical adjacency matrix $\mathbf{A}^{\text{geo}} \in \mathbb{R}^{N \times N}$, where $a_{i,j}^{\text{geo}} = 1$ indicates that regions $i$ and $j$ are geographically connected.

\subsection{EpiGNN Model Framework}
EpiGNN is an epidemic forecasting framework that integrates spatio-temporal modeling to capture geographic and temporal patterns, with a learnable graph structure that dynamically infers transmission relationships. It takes historical regional case data through multi-scale convolutions for analyzing trends at different time spans, a transmission risk encoder incorporating mobility and environmental factors, RAGL,and GCN for propagation modeling \cite{xie2022epignn}. Its process is as follows:

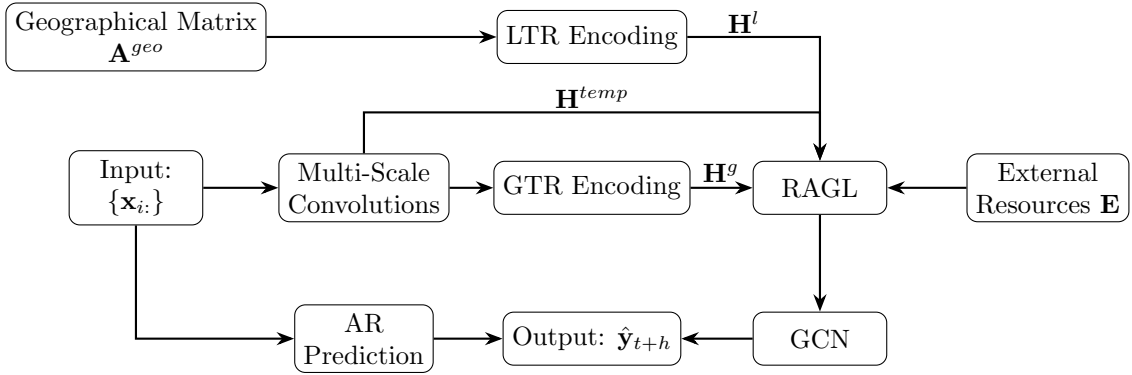
\begin{figure}
        \centering
        \begin{tikzpicture}[
            box/.style={rectangle, draw, rounded corners, minimum height=2em, minimum width=5em, align=center},
            arrow/.style={-Stealth, thick}
        ]
            \node[box] (input) at (0,0) {Input:\\ $\mathbf\{\mathbf{x}_{i:}\}$};
            \node[box] (geo) at (0,2) {Geographical Matrix \\ $\mathbf{A}^{geo}$};
            \node[box] (conv) at (3,0) {Multi-Scale\\Convolutions};
            \node[box] (gtr) at (6,0) {GTR Encoding};
            \node[box] (ltr) at (6,2) {LTR Encoding};
            \node[box] (ragl) at (9,0) {RAGL};
            \node[box] (ER) at (12,0) {External\\Resources $\mathbf{E}$};
            \node[box] (gcn) at (9,-2) {GCN};
            \node[box] (ar) at (3,-2) {AR\\Prediction};
            \node[box] (output) at (6,-2) {Output: $\hat{\mathbf{y}}_{t+h}$};
            \node (htemp) at (6,1.2) {$\mathbf{H}^{temp}$};
            \node (hl) at (8,2.2) {$\mathbf{H}^{l}$};
            \node (hg) at (7.7,0.2) {$\mathbf{H}^{g}$};
            
            \draw[arrow] (geo) -- (ltr);
            \draw[arrow] (input) -- (conv);
            \draw[arrow] (conv) -- (gtr);
            \draw[arrow] (conv) |- ++(1.5,1) -| (ragl); 
            \draw[arrow] (ltr) -| (ragl);
            \draw[arrow] (gtr) -- (ragl);
            \draw[arrow] (ER) -- (ragl);
            \draw[arrow] (ragl) -- (gcn);
            \draw[arrow] (gcn) -- (output);
            \draw[arrow] (input) |- (ar);
            \draw[arrow] (ar) -- (output);
        \end{tikzpicture}
        \caption{EpiGNN workflow}
\end{figure}

\begin{itemize}

    \item \textbf{Multi-Scale Temporal Feature Extraction (Region-Aware Convolution)}\\
    This module extracts rich temporal features for each region $i$ from its time series $\mathbf{x}_{i:} \in \mathbb{R}^{1 \times T}$ using a multi-branch convolutional structure. It consists of several parallel convolutional branches, each designed to capture distinct temporal patterns by employing specific combinations of kernel sizes and dilation rates. These branches are categorized as follows:
    \begin{itemize}
        \item \textbf{Local Pattern Branches ($M_L$ branches):} These branches utilize relatively small kernel sizes $s_l$ and a dilation rate of $r_l=1$ to capture consecutive short-term dependencies.
        \item \textbf{Periodic Pattern Branches ($M_P$ branches):} These branches also use relatively small kernel sizes $s_p$ but employ larger dilation rates $r_p > 1$ to capture recurrent or periodic short-term trends that span across intervals.
        \item \textbf{Global Pattern Branches ($M_G$ branches):} These branches employ a kernel size equal to the full time window length $T$ (with a dilation rate of $r_g=1$) to capture comprehensive, overarching patterns across the entire observation period.
    \end{itemize}
    The total number of branches is $M = M_L + M_P + M_G$. Each of the $M$ branches employs $k$ filters and all filters within a given branch use the same kernel size and dilation rate. For a convolutional filter $\mathbf{f}_{n, 1 \times s, r} \in \mathbb{R}^{1 \times s}$ (where $n=1, \ldots, k$ is the filter index), it represents a kernel of size $s$ applied along the temporal dimension, with a width of 1, and a temporal dilation rate $r$. The convolution operation at step $j$ for location $i$ is defined as:
    \begin{equation}
        \left(\mathbf{x}_{i:} \textcolor{black}{\star} \mathbf{f}_{n,1 \times s,r}\right)[j] = \sum_{p=0}^{s-1} \mathbf{f}_{n,1 \times s,r}[0, p] \cdot \mathbf{x}_{i:}[j - r \times p]
    \end{equation}
    Here, $\textcolor{black}{\star}$ is convolution operation. Let $\mathbf{X}^{(m)}_{i} \in \mathbb{R}^{k \times L_m}$ denote the output feature maps for region $i$ from the $m$-th convolutional branch, where $L_m$ is the uniform output length of the feature maps produced by all filters within that branch. After most convolutional operations (specifically, for local and periodic pattern branches), an adaptive max pooling layer, denoted as $\text{Pool}(\cdot)$, is applied to compress the feature maps along the time axis to a fixed dimension $P_u$. This helps in controlling the feature size and extracting salient information. Global pattern branches, having already aggregated information across the entire window by design ($L_m=1$), are not subjected to further pooling.
    
    All branch outputs are then flattened along their temporal dimension and concatenated. The concatenated feature vector is subsequently passed through a non-linear activation function, $\text{Activation}(\cdot)$ (e.g., hyperbolic tangent), to form the final temporal feature embedding for region $i$:
    \begin{equation}
        \mathbf{h}_i^{temp} = \text{Activation}\left( \text{Concat}_{m=1}^{M} \left[ \text{Flatten}(\text{Pool}_m(\mathbf{X}^{(m)}_{i})) \right] \right)
    \end{equation}
    where $\text{Pool}_m$ applies pooling if $L_m > 1$. The resulting temporal feature embedding for each region is $\mathbf{h}_i^{temp} \in \mathbb{R}^D$. The total feature dimension $D$ for a single region is the sum of the flattened dimensions from all $M$ branches. That is, $D = \sum_{m=1}^{M} (k \times P'_m)$, where $P'_m$ is $P_u$ for pooled branches, and $1$ for non-pooled branches. The specific number of branches for each category ($M_L, M_P, M_G$), their respective kernel sizes ($s_l, s_p, s_g$), dilation rates ($r_l, r_p, r_g$), filter count $k$, and pooling dimensions $P_u$ are hyperparameters determined experimentally to best capture the dataset's characteristics, such as periodicity in epidemic data.

    \item \textbf{Local Transmission Risk (LTR) Encoding ($\mathbf{A}_{i:}^{geo}$)}\\
    This module quantifies the local transmission risk for each region based on its geographical connectivity. It leverages a geographical adjacency matrix $\mathbf{A}^{geo}$, where $a_{i,j}^{geo}=1$ signifies that region $i$ and region $j$ are direct neighbors (with $a_{i,i}^{geo}=1$ by default for self-connectivity).

    For each region $i$, its local transmission risk is captured by a vector $\mathbf{h}_i^l \in \mathbb{R}^D$. This encoding is derived from the region's degree, $d_i$, which is calculated as the sum of elements in the $i$-th row of the adjacency matrix, $\mathbf{A}_{i:}^{geo}$, i.e., $d_i = \sum_j a_{i,j}^{geo}$. This degree represents the total count of region $i$'s geographical neighbors. The rationale is that a higher degree indicates greater connectivity and centrality, leading to more frequent potential interactions and an elevated likelihood of disease spread.
    
    The local transmission risk encoding is formulated as:
    \begin{equation}
        \mathbf{h}_i^l = \mathbf{W}^l \cdot d_i + \mathbf{b}^l
    \end{equation}
    
    where $\mathbf{W}^l\in \mathbb{R}^D$ is a learnable weight vector that adjusts the importance of this degree, allowing the model to learn how connectivity translates into a multi-dimensional risk representation. $\mathbf{b}^l\in \mathbb{R}^D$ is a learnable bias vector, also optimized during training, which adjusts the baseline risk for each dimension.

    \item \textbf{Global Transmission Risk (GTR) Encoding ($\mathbf{H}^{temp}$)}\\
    This module evaluates the risk of disease transmission across all regions from a global perspective, capturing potential infection pathways spread beyond local neighbors.These pathways include non-adjacent connections (e.g., regions linked by long-distance travel despite lacking shared borders), cross-regional interactions (e.g., disease spread between distant administrative or geopolitical areas), and periodic patterns (e.g., seasonal fluctuations driven by climate or recurring travel cycles). Following the idea of multiplicative self-attention\cite{vaswani2017attention, Sukhbaatar2015}, first, $\mathbf{H}^{temp} = [\mathbf{h}_1^{temp}, \mathbf{h}_2^{temp}, \ldots, \mathbf{h}_N^{temp}]^T \in \mathbb{R}^{N \times D}$ is projected into two separate embedding spaces via learned linear transformations $\mathbf{W}^q\in \mathbb{R}^{D\times F}$ and $\mathbf{W}^k \in \mathbb{R}^{D\times F}$ to generate the Query matrix $\mathbf{Q}\in \mathbb{R}^{N\times F}$ and the Key matrix $\mathbf{K}\in \mathbb{R}^{N\times F}$. Here, $F$ denotes the attention projection dimension, representing the size of the latent feature space to which the original input vectors are mapped for computing pairwise attention scores. These matrices allow the model to represent each region in two roles: as a potential receiver (Query) and as a potential transmitter (Key) of infection risk. The Attention Scores matrix $\mathbf{A} \in \mathbb{R}^{N \times N}$ is computed as the dot product between $\mathbf{Q}$ and $\mathbf{K}^\text{T}$, quantifying correlations between all pairs of positions: 
    \begin{equation}
        \mathbf{Q} = \mathbf{H}^{temp}\mathbf{W}^q,
        \mathbf{K} = \mathbf{H}^{temp}\mathbf{W}^k,
        \mathbf{A} = \mathbf{Q}\mathbf{K}^\text{T}
    \end{equation}
    After being normalized, $a_{i,j}$ represents the transmission risk of region $j$ to region $i$:
    \begin{equation}
        a_{i,j} = \frac{a_{i,j}}{\max(||\mathbf{a}_{i:}||_2, \epsilon)}
    \end{equation}
    where $\mathbf{a}_{i:} \in \mathbb{R}^N$ is the $i$-th row of matrix $\mathbf{A}$, $||\cdot||_2$ is the $L_2$ norm and $\epsilon$ is a small value to prevent division by zero, ensuring numerical stability. By summing all incoming risks $a_{i,j}$ from other regions $j$, $g_i$ quantifies the aggregated transmission risk of all other regions to region $i$ :
    \begin{equation}
        g_i = \sum_j a_{i,j}
    \end{equation}
    The aggregated risk score $g_i$ is then linearly transformed to produce the final global transmission risk vector $\mathbf{h}_i^g \in \mathbb{R}^D$ for region $i$:
    \begin{equation}
        \mathbf{h}_i^g = \mathbf{W}^g \cdot g_i + \mathbf{b}^g
    \end{equation}
    where $\mathbf{W}^g \in \mathbb{R}^D $ and  $\mathbf{b}^g \in \mathbb{R}^D $  are trainable parameters, randomly initialized and optimized during training.

    \item \textbf{Region-Aware Graph Learner ($\mathbf{h}_{i}^{temp},\mathbf{h}_{j}^{temp},\mathbf{A}^{geo},\mathbf{E}$)}\\
    First, the model learns an asymmetric temporal correlation graph $\hat{\mathbf{A}}$ from time series data, capturing directed temporal influence between regions. Next, the geographical adjacency matrix $\mathbf{A}^{geo}$ and the region degree information $\mathbf{d}=[d_1,\ldots,d_N]^T\in \mathbb{R}^N$ are used as gates to modulate spatial dependencies, resulting in a spatial dependency-aware correlation matrix $\mathbf{A}^{spa}$. If external resources (e.g., human mobility data or environmental factors) are available, they are encoded as an external correlation matrix $\mathbf{A}^e$, further enriching the model’s understanding of true inter-region interactions. The final propagation graph is constructed by summing $\hat{\mathbf{A}}$, $\mathbf{A}^{spa}$ and $\mathbf{A}^e$, forming the ultimate propagation graph structure. Using broadcasting, dynamic temporal relationships are extracted by:
    \begin{equation}
        \mathbf{M}_1 = \tanh(\mathbf{H}^{temp}\mathbf{W}_1 + \mathbf{b}_1), \quad \mathbf{M}_2 = \tanh(\mathbf{H}^{temp}\mathbf{W}_2 + \mathbf{b}_2) \in \mathbb{R}^{N \times F}
    \end{equation}
    \begin{equation}
        \hat{\mathbf{A}} = \text{ReLU}(\tanh(\mathbf{M}_1\mathbf{M}_2^\text{T} - \mathbf{M}_2\mathbf{M}_1^\text{T}))  \in \mathbb{R}^{N \times N}
    \end{equation}
    where $\mathbf{W_1},\mathbf{W_2} \in \mathbb{R}^{D \times F}$ are weight matrices and $\mathbf{b}_1,\mathbf{b}_2\in \mathbb{R}^{1 \times F}$ are bias vectors. Spatial dependency combines the geographical adjacency matrix with degree gating:
    \begin{equation}
        \mathbf{D}^* = \text{sigmoid}(\mathbf{W}^{*} \odot \mathbf{d}\mathbf{d}^\text{T})  \in \mathbb{R}^{N \times N}
    \end{equation}
    \begin{equation}
        \mathbf{A}^{spa} = \mathbf{D}^* \odot \mathbf{A}^{geo}  \in \mathbb{R}^{N \times N}
    \end{equation}

    where $\odot$ denotes element-wise multiplication and $\mathbf{W^{*}}\in \mathbb{R}^{N\times N}$ is the learnable parameter matrix. After fusing external resources (e.g., human mobility data $\mathbf{E}=[\mathbf{E_1},\mathbf{E_2},\ldots,\mathbf{E_t}]$, where $\mathbf{E}_z \in \mathbb{R}^{N \times N}$ is the correlation matrix at time step z):
    \begin{equation}
        \mathbf{A}^e = \mathbf{W}^e \odot \sum_{i=0}^{e-1} \mathbf{E}_{t-i}  \in \mathbb{R}^{N \times N}
    \end{equation}
    \begin{equation}
        \tilde{\mathbf{A}}=\hat{\mathbf{A}}+ \mathbf{A}^{spa} + \mathbf{A}^e 
    \end{equation}
    where $e$ is the look-back window of external resources, and $\mathbf{W}^e \in \mathbb{R}^{N \times N}$ is a learnable matrix.
    \item \textbf{Graph Convolution Network ($\mathbf{h}_i^{feat},\tilde{\mathbf{A}}$)}\\
    The temporal feature $\mathbf{h}_i^{temp}$, local transmission risk encoding $\mathbf{h}_i^l$, and global transmission risk encoding $\mathbf{h}_i^g$ for region $i$ are summed to obtain $\mathbf{h}_i^{feat}$, which serves as the initial representation for each region node.
    \begin{equation}
        \mathbf{h}_i^{feat} = \mathbf{h}_i^{temp} + \mathbf{h}_i^l + \mathbf{h}_i^g \in \mathbb{R}^{D}
    \end{equation}
    
    Using the propagation graph structure $\tilde{\mathbf{A}}$ constructed in the previous stage, the node representation for each layer is updated as follows to obtain the node representation $\mathbf{h}_i^{(l)}$ for region $i$ at layer $l$, providing structured spatio-temporal feature input for the final prediction layer.
    \begin{equation}
        \mathbf{H}^{(l)} = \sigma(\tilde{\mathbf{D}}^{-1}\tilde{\mathbf{A}}\mathbf{H}^{(l-1)}\mathbf{W}^{(l-1)}) \in \mathbb{R}^{N \times D}
    \end{equation}
    
    where $\mathbf{H}^{(0)} = \mathbf{H}^{feat} = [\mathbf{h}_1^{feat},\mathbf{h}_2^{feat},\ldots,\mathbf{h}_N^{feat}]^T\in \mathbb{R}^{N \times D}$, $\mathbf{W}^{(l)}\in \mathbb{R}^{D \times D}$ is the weight matrix,  $\tilde{\mathbf{D}}\in \mathbb{R}^{N \times N}$ is the degree matrix of  $\tilde{\mathbf{A}}$,  and $\sigma$ is a non-linear activation function (e.g., ELU).

    \item \textbf{Output ($\mathbf{x}_{i:},[\mathbf{h}_i^{feat};\mathbf{h}_i^{(L)}]$)}\\
    Finally, the GCN input $\mathbf{h}_i^{feat}$ is concatenated with the last layer's output $\mathbf{h}_i^{(L)}$ ($L$ denotes the last layer), passed through a fully connected layer to obtain the non-linear prediction result $\hat{y}^{nl}_{i,t+h}$.
    \begin{equation}
        \hat{\mathbf{y}}_{t+h}^{nl} = [\mathbf{H}^{(0)}; \mathbf{H}^{(L)}]\mathbf{W}^{nl} + \mathbf{b}^{nl} \in \mathbb{R}^N
    \end{equation}
    
    where $[\cdot;\cdot]$ denotes feature concatenation. $\mathbf{W}^{nl} \in \mathbb{R}^{2D}$ and $\mathbf{b}^{nl} \in \mathbb{R}^N$ are the parameters. The linear prediction result from the AR module (optional) $\hat{\mathbf{y}}_{t+h}^{ar} \in \mathbb{R}^N$ and the non-linear prediction result from the GCN output $\hat{\mathbf{y}}_{t+h}^{nl}$ are summed to obtain the final predicted value $\hat{\mathbf{y}}_{t+h}$.
    \begin{equation}
        \hat{y}_{i,t+h}^{ar} = \sum_{\beta =0}^{q-1} \mathbf{W}_\beta ^{ar} \cdot x_{i,t-\beta } + b^{ar}
    \end{equation}
    \begin{equation}
        \hat{\mathbf{y}}_{t+h} = \hat{\mathbf{y}}_{t+h}^{ar} + \hat{\mathbf{y}}_{t+h}^{nl}
    \end{equation}
    where $\mathbf{W}^{ar} \in \mathbb{R}^q$ and $b^{ar} \in \mathbb{R}$ are the parameters in AR component, and $q$ is the look-back window of AR.
    The loss function is the mean squared error:
    \begin{equation}
        Loss(\Theta) = ||\mathbf{y}_{t+h} - \hat{\mathbf{y}}_{t+h}||_2^2
    \end{equation}
    where $\Theta$ represents the set of all trainable parameters, $\mathbf{y}_{t+h}$ is the ground truth value, and $||\cdot||_2^2$ is squared $L_2$ norm.
\end{itemize}

\subsection{ColaGNN Model Framework}
ColaGNN is a class of Graph Neural Network models specifically designed for long-term influenza forecasting. Its main features include a temporal dependency modeling, a cross-location attention mechanism and a graph-based feature propagation mechanism, aiming to enhance the ability to model complex long-term trends and dependencies between non-adjacent regions\cite{deng2020cola}. Its overall structure is as follows :

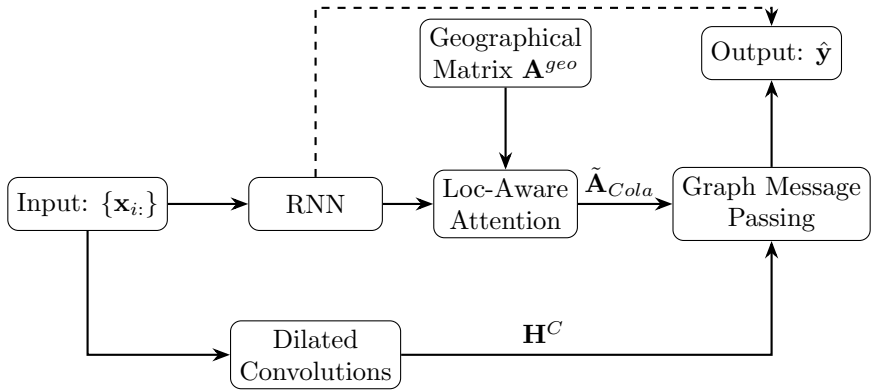
\begin{figure}
        \centering
        \begin{tikzpicture}[
            box/.style={rectangle, draw, rounded corners, minimum height=2em, minimum width=5em, align=center},
            arrow/.style={-Stealth, thick}
        ]
            \node[box] (input) at (0,0) {Input: $\{\mathbf{x}_{i:}\}$};
            \node[box] (rnn) at (3,0) {RNN};
            \node[box] (conv) at (3,-2) {Dilated\\Convolutions};
            \node[box] (attn) at (5.5,0) {Loc-Aware\\Attention};
            \node[box] (gcn) at (9,0) {Graph Message\\Passing};
            \node[box] (output) at (9,2) {Output: $\hat{\mathbf{y}}$};
            \node(hc) at (6,-1.7) {$\mathbf{H}^C$};
            \node(a) at (7,0.3) {$\tilde{\mathbf{A}}_{Cola}$};
            \node[box] (geo) at (5.5,2) {Geographical\\ Matrix $\mathbf{A}^{geo}$};

            \draw[arrow] (geo) -- (attn);
            \draw[arrow] (input) -- (rnn);
            \draw[arrow] (input) |- (conv);
            \draw[arrow, dashed] (rnn) |- ++(6,2.6) -|(output);
            \draw[arrow] (attn) -- (gcn);
            \draw[arrow] (conv) -| (gcn);
            \draw[arrow] (gcn) -- (output);
            \draw[arrow] (rnn) -- (attn);
        \end{tikzpicture}
        \caption{ColaGNN workflow}
    \end{figure}

\begin{itemize}
\item\textbf{Temporal Dependency Modeling Module}

This module integrates two complementary approaches to capture diverse temporal information from the historical time series of each region: RNN for long-term sequence aggregation and CNN for multi-scale local pattern extraction.

\begin{itemize}
    \item \textbf{Long-term Sequence Aggregation: RNN Module ($\mathbf{x}_{i:}$)}\\
    The purpose of this module is to capture comprehensive long-term dependencies and sequential information from the entire historical time series of each region, treating the data as a continuous sequence. For each region $i$, its historical input sequence $\mathbf{x}_{i:} = [x_{i,t-T+1}, ..., x_{i,t}]$ is fed into an RNN model (e.g., GRU \cite{Cho2014} or LSTM \cite{Hochreiter1997}) with shared parameters across regions. The RNN aggregates and learns all features of historical time steps equally, updating its hidden state $\mathbf{h}_{i,t^{\prime}}$ at each time step. The final hidden state $\mathbf{h}_{i,t}$ serves as the holistic temporal representation for region $i$. Let $D^{\prime}$ be the dimension of the hidden state. The RNN updates the hidden state at time $t^{\prime}$:
\begin{equation}
        \mathbf{h}_{i,t^{\prime}} = \tanh(\mathbf{w}x_{i,t^{\prime}} + \mathbf{U}\mathbf{h}_{i,t^{\prime}-1} + \mathbf{b}) \in \mathbb{R}^{D^{\prime}}
\end{equation}

where $\mathbf{h}_{i,t^{\prime}}$ and $\mathbf{h}_{i,t^{\prime}-1}$ are the hidden states at time $t^{\prime}$ and $t^{\prime}-1$, respectively, $\tanh$ is the non-linear activation function, $\mathbf{w} \in \mathbb{R}^{D^{\prime}}$, $\mathbf{U} \in \mathbb{R}^{{D^{\prime}} \times {D^{\prime}}}$, $\mathbf{b} \in \mathbb{R}^{D^{\prime}}$ are trainable weights and biases. The last hidden state $\mathbf{h}_{i,t}$ serves as the representation for location $i$, used for subsequent attention calculation. 

\item \textbf{Local Pattern Extraction: Multi-Scale
Convolutional Module (ColaGNN's)}\\
This module is designed to extract important local and multi-scale feature patterns from the time series $\mathbf{x}_{i:} \in \mathbb{R}^{1 \times T}$ for region $i$ using two parallel one-dimensional convolutional branches: one for capturing short-term patterns and another for long-term patterns. Each branch employs $k^{\prime}$ filters. For a filter $\mathbf{c}^{(r',n')} \in \mathbb{R}^{s'}$ (where $n'=1, \ldots, k^{\prime
}$ is the filter index, $s'$ is the kernel size, and $r'$ is the dilation rate), the convolution operation at step $j$ for location $i$ is:
\begin{equation}
    \mathbf{d}_i^{(r',n')}[j] = \sum_{p=0}^{s'-1} \mathbf{x}_{i:}[j + r' \times p] \cdot \mathbf{c}^{(r',n')}[p]
\end{equation}
Here, $\mathbf{d}_i^{(r',n')}$ represents the output feature vector for the $n'_{th}$ filter with dilation rate $r'$. The output length of a 1D convolution $L_{out}$ given input length $T$, kernel size $s'$, and dilation rate $r'$ (with stride 1 and padding 0) is calculated as $L_{out} = T - r' \times (s' - 1)$. The two convolutional branches are configured with distinct kernel sizes and dilation rates:
\begin{itemize}
    \item \textbf{Short-term Pattern Branch:} This branch is designed to capture immediate, fine-grained temporal dependencies or overall trends spanning the entire input window. It achieves this by utilizing a relatively large kernel size $s'_{short}$ (e.g., matching the input length $T$) and a small dilation rate $r'_{short}$ (typically $1$). For each of its $k^{\prime}$ filters, this branch generates an output feature map $\mathbf{d}_i^{(r'_{short},n')} \in \mathbb{R}^{L_{short}}$, where $L_{short} = T - r'_{short} \times (s'_{short} - 1)$.
    \item \textbf{Long-term Pattern Branch:} This branch focuses on capturing more distant, sparse, or periodic dependencies over a broader temporal range. It achieves this by employing a relatively smaller kernel size $s'_{long}$ but a larger dilation rate $r'_{long}$ ($r'_{long} > r'_{short}$). The larger dilation rate effectively expands the receptive field without increasing the number of parameters, allowing the filter to skip intermediate time steps and focus on distant patterns. For each of its $k^{\prime}$ filters, this branch generates an output feature map $\mathbf{d}_i^{(r'_{long},n')} \in \mathbb{R}^{L_{long}}$, where $L_{long} = T - r'_{long} \times (s'_{long} - 1)$.

\end{itemize}
The outputs from all filters across both branches are flattened and concatenated. This combined feature vector is then passed through a Rectified Linear Unit (ReLU) activation function to form the final temporal feature embedding $\mathbf{h}_i^C$:
\begin{equation}
    \mathbf{h}_i^C = \text{ReLU}\left\{ \text{Concat}\left[ \mathbf{d}_i^{(r'_{short},1)}, \ldots, \mathbf{d}_i^{(r'_{short},k^{\prime})}, \mathbf{d}_i^{(r'_{long},1)}, \ldots, \mathbf{d}_i^{(r'_{long},k^{\prime})} \right] \right\}
\end{equation}
The concatenated outputs form $\mathbf{h}_i^C \in \mathbb{R}^{F^{(0)}}$. The total feature dimension $F^{(0)}$ for a single region $i$ after these multi-scale convolutions and concatenation is calculated as:
\begin{equation}
    F^{(0)} = k^{\prime} \times (L_{short} + L_{long})
\end{equation}
The specific kernel sizes ($s'_{short}, s'_{long}$), dilation rates ($r'_{short}, r'_{long}$), and number of filters $k^{\prime}$ are hyperparameters determined experimentally to best capture the dataset's characteristics, such as periodicity in epidemic data.

\end{itemize}

\item\textbf{Cross-Regional Influence Modeling Module: Loc-Aware Attention ($h_{i,t},h_{j,t},\mathbf{A}^{geo}$)}\\
The purpose of this module is to capture dynamic propagation relationships between regions, not limited to fixed geographical adjacency. For each pair of regions $(i,j)$, an asymmetric attention weight $a_{i,j}^{\prime}$ is calculated based on their historical hidden states $h_{i,t}$ and $h_{j,t}$, combined with the geographical adjacency matrix $\mathbf{A}^{geo}$, forming a location-aware dynamic adjacency matrix. This module introduces a gating mechanism to fuse static geographical adjacency with dynamic propagation trends.

\begin{itemize}
    \item\textbf{General Attention Calculation Mechanism:}\\
Based on RNN hidden states, additive attention \cite{Bahdanau2014} is used to calculate the influence of location $j$ on location $i$:
    \begin{equation}
        a_{i,j}^{\prime} = \mathbf{v}^\text{T} \Delta (\mathbf{W}^{\prime}\mathbf{h}_{i,t} + \mathbf{W}^{\prime\prime} \mathbf{h}_{j,t} + \mathbf{b}^{\prime}) + b^v
    \end{equation}

where $\Delta $ is an element-wise activation function (e.g., ELU), $\mathbf{W}^{\prime}, \mathbf{W}^{\prime\prime} \in \mathbb{R}^{d_a \times {D^\prime}}$ are weight matrices, $d_a$ is a hyperparameter, $\mathbf{v} \in \mathbb{R}^{d_a}, \mathbf{b}^{\prime} \in \mathbb{R}^{d_a}, b^v \in \mathbb{R}$ are trainable parameters, and $a_{i,j}^{\prime}$ is the attention coefficient, representing the influence of $j$ on $i$. The resulting asymmetric attention matrix $\mathbf{A^{\prime}} \in \mathbb{R}^{N \times N}$ has each row representing the influence of other locations on the current location. To normalize the overall influence of different locations, $L_{p^{\prime}}$ norm normalization is applied row-wise:
    \begin{equation}
        \mathbf{a}_{i:}^{\prime} \leftarrow \frac{\mathbf{a}_{i:}^{\prime}}{\max(||\mathbf{a}_{i:}^{\prime}||_{p^{\prime}}, \epsilon)}
    \end{equation}

where $||\cdot||_{p^{\prime}}$ is the $L_{p^{\prime}}$ norm (in experiments, $p^{\prime}=2$). 

\item\textbf{Location-Aware Fusion Mechanism (Gate):}\\
To enhance geographical interpretability, the static geographical adjacency matrix $\mathbf{A}^{geo}$ is introduced and normalized, allowing the model to consider both data-driven correlations and existing geographical information without excessive deviation. The attention results are weighted and fused with the original geographical adjacency matrix to generate a location-aware attention matrix $\tilde{\mathbf{A}}_{Cola}$, preserving both historical dynamics and geographical priors.
    \begin{equation}
        \tilde{\mathbf{A}}^{geo} = \mathbf{D}^{-\frac{1}{2}}\mathbf{A}^{geo}\mathbf{D}^{-\frac{1}{2}} \in \mathbb{R}^{N \times N}
    \end{equation}
where $\mathbf{D} \in \mathbb{R}^{N \times N}$ is the degree matrix of $\mathbf{A}^{geo}$ defined as $d_{ii} = \sum_{j=1}^N a_{ij}^{geo}$.
The gating matrix $\mathbf{M}$ is learned based on the general attention matrix $\mathbf{A}^{\prime}$:
    \begin{equation}
        \mathbf{M} = \text{Sigmoid}(\mathbf{W}^M \mathbf{A^{\prime}} + b^M \mathbf{1}_N \mathbf{1}_N^\text{T})
    \end{equation}

where $\mathbf{W}^M \in \mathbb{R}^{N \times N}, b^M \in \mathbb{R}$ are trainable parameters, and $\mathbf{1}_N \in \mathbb{R}^N$ is an $N$-dimensional vector whose elements are all $1_s$.
The final location-aware attention matrix is:
    \begin{equation}
        \tilde{\mathbf{A}}_{Cola} = \mathbf{M} \odot \tilde{\mathbf{A}}^{geo} + (\mathbf{1}_N \mathbf{1}_N^\text{T} - \mathbf{M}) \odot \mathbf{A}^{\prime} \in \mathbb{R}^{N \times N}
    \end{equation}

$\tilde{\mathbf{A}}_{Cola}$ dynamically balances geographical adjacency and historical data-driven attention, enhancing the model's ability to capture non-adjacent regional dependencies.
\end{itemize}

\item\textbf{Graph Message Passing ($\mathbf{h}_i^C,\tilde{\mathbf{A}}_{Cola}$)}\\
The purpose of this module is to integrate neighborhood information and model potential correlated propagation patterns between regions. The location-aware adjacency matrix $\tilde{\mathbf{A}}_{Cola}$ generated in the previous step serves as the graph structure, and temporal features are used as node features for graph neural network propagation. Dilated convolutional features $\mathbf{h}_i^C$ act as initial node features $\hat{\mathbf{h}}_i^{(0)}$, performing graph message passing where each node receives (weighted) information updates from neighbors to obtain a comprehensive representation that incorporates the influence of surrounding regions:
    \begin{equation}
        \hat{\mathbf{h}}_i^{(l)} = \text{ReLU}\left(\sum_{j \in \mathscr{N}} a^{\prime\prime}_{i,j} \hat{\mathbf{W}}^{(l-1)}\hat{\mathbf{h}}_j^{(l-1)} + \mathbf{b}^{(l-1)}\right)
    \end{equation}

where $\hat{\mathbf{h}}_j^{(l-1)} \in \mathbb{R}^{F^{(l-1)}}$ is node feature of node $j$ at layer $l-1$, $a^{\prime\prime}_{i,j}$ is the $(i,j)$ element of $\tilde{\mathbf{A}}_{Cola}$, $\hat{\mathbf{W}}^{(l-1)} \in \mathbb{R}^{F^{(l)} \times F^{(l-1)}}$ is the weight matrix for hidden layer $l$ with $F^{(l)}$ feature maps, $\mathbf{b}^{(l-1)} \in \mathbb{R}^{F^{(l)}}$ is the bias, and $\mathscr{N}$ is the set of locations. Multi-layer message passing (2 layers in experiments) updates node representations, capturing inter-regional propagation dynamics.

\item\textbf{Output ($[\mathbf{h}_{i,t}; \hat{\mathbf{h}}_i^{(L^{\prime})}]$)}\\
The result after graph propagation $\hat{\mathbf{h}}_i^{(L^{\prime})}$ ($L^{\prime}$ denotes the last layer) is concatenated with the RNN hidden state $\mathbf{h}_{i,t}$ and used as the final input. A linear layer outputs the prediction of future values:
    \begin{equation}
       \hat{y}_i = \mathbf{W}^T_o [\mathbf{h}_{i,t}; \hat{\mathbf{h}}_i^{(L^{\prime})}]+ b_o
    \end{equation}

where $\mathbf{W}_o \in \mathbb{R}^{D^{\prime}+F^{(L^{\prime})}}, b_o \in \mathbb{R}$ are trainable parameters.

\item\textbf{Loss Function}\\
To prevent overfitting in small-sample scenarios, ColaGNN additionally incorporates an $L2$ regularization term into the loss function, which penalizes large model parameters, ensuring sparsity and robustness. The complete loss function is as follows:
    \begin{equation}
       Loss(\Theta^{\prime}) = \sum_{i=1}^N \sum_{m_{i}=1}^{n_i} |y_{i,m_{i}} - \hat{y}_{i,m_{i}}|  + \lambda \Re(\Theta^{\prime})
    \end{equation}
where $n_i$ is the number of samples in location $i$ obtained by a moving window, shared by all locations, $y_{i,m_{i}}$ is the true value of location $i$ in sample $m_{i}$, and $\hat{y}_{i,m_{i}}$ is the model prediction. $\Re(\Theta^{\prime})$ is the regularization term, and $\Theta^{\prime}$ represents the set of all trainable parameters in the model. The impact of the regularization term on the total loss can be adjusted by controlling the magnitude of the regularization coefficient $\lambda$.
\end{itemize}

\section{Experiments and Analysis}
Below, we will conduct experiments on EpiGNN and ColaGNN models using four different datasets and compare their prediction performance under various metrics.

\subsection{Experimental Setup}

\begin{itemize}
    \item\textbf{Datasets:} Three influenza datasets (Japan-Prefectures, US-Regions, US-States) and one COVID-19 dataset (Australia-COVID). Data is chronologically divided into training (50\%), validation (20\%), and testing (30\%) sets.
    \begin{itemize}
        \item\textbf{Australia-COVID:} This dataset is obtained from the publicly available JHU-CSSE dataset \cite{australia-covid}, this dataset records daily new COVID-19 confirmed cases in Australia, encompassing 6 states and 2 territories, from January 27, 2020, to August 4, 2021.
        \item\textbf{Japan-Prefectures:} This dataset is sourced from the Infectious Diseases Weekly Report (IDWR) in Japan \cite{japan-ili}, providing weekly influenza-like-illness (ILI) statistics (patient counts) across 47 prefectures from August 2012 to March 2019.
        \item\textbf{US-Regions:} This dataset is derived from the ILINet section of the US-HHS dataset \cite{us-hhs}, including weekly influenza activity levels for 10 Health and Human Services (HHS) regions across the U.S. mainland, covering the period from 2002 to 2017. Each region aggregates flu patient counts, calculated by combining state-specific data.
        \item\textbf{US-States:} This dataset is collected from the Center for Disease Control (CDC) \cite{us-cdc}, containing weekly counts of patient visits for ILI (positive cases) for each state in the United States from 2010 to 2017. After excluding Florida due to missing data, the dataset retains 49 states.
    \end{itemize}

    \begin{table}[H]
    \centering
    \setlength{\tabcolsep}{4pt} 
    \resizebox{\textwidth}{!}{
    \begin{tabular}{@{} 
      l 
      S[table-format=2.0] 
      S[table-format=3.0] 
      *{4}{S[table-format=5.0]} 
      l @{}
    }
    \toprule
    \textbf{Datasets} & {\textbf{Regions}} & {\textbf{Length}} & {\textbf{Min}} & {\textbf{Max}} & {\textbf{Mean}} & {\textbf{SD}} & \textbf{Gran.} \\
    \midrule
    Japan-Prefectures & 47 & 348 & 0 & 26635 & 655 & 1711 & weekly \\
    US-Regions & 10 & 785 & 0 & 16526 & 1009 & 1351 & weekly \\
    US-States & 49 & 360 & 0 & 9716 & 223 & 428 & weekly \\
    Australia-COVID & 8 & 556 & 0 & 9987 & 539 & 1532 & daily \\
    \bottomrule
    \end{tabular}
    }
    \caption{Dataset characteristics summary}
    \label{tab:datasets}
    \end{table}

\item\textbf{Baseline:} STGCN

\item\textbf{Evaluation Metrics:} The models use RMSE, MAE, and PCC as evaluation metrics.
\[
\text{RMSE} = \sqrt{\frac{1}{n}\sum_{i=1}^n (\hat{y}_i - y_i)^2}
\]
\[
\text{MAE} = \frac{1}{n}\sum_{i=1}^n |\hat{y}_i - y_i|
\]
\[
\text{PCC} = \frac{\sum_{i=1}^n (\hat{y}_i - \overline{\hat{y}})(y_i - \overline{y})}{\sqrt{\sum_{i=1}^n (\hat{y}_i - \overline{\hat{y}})^2}\sqrt{\sum_{i=1}^n (y_i - \overline{y})^2}}
\]

\item\textbf{Hyperparameters:} Look-back window $T=20$, prediction horizons $h=\{2,5,8,11,14,17,\\20,23,26,29,32\}$, batch size 128. To improve training stability and convergence speed, the Adam optimizer \cite{Kingma2015} (Adaptive Moment Estimation) is introduced to dynamically adjust the learning rate: initial learning rate $0.001$, weight decay $5e-4$. To further enhance the model's generalization ability, Dropout and Early Stopping are incorporated during training: Dropout = $0.2$, Early Stopping = 100.
\item\textbf{Environment:} 
Our experiments were conducted on a Windows server equipped with an Nvidia 2060 GPU, leveraging CUDA 11.8 for GPU acceleration. The core computational framework utilized was PyTorch 2.2.1. The following key Python packages were instrumental in our implementation and analysis:
    \begin{itemize}
    \item\textbf{pandas (version 2.2.3):} 
     Employed for efficient data manipulation and analysis.
     \item\textbf{numpy (version 1.26.4):}
    Used for fundamental numerical operations and array processing.
    \item\textbf{scikit-learn (version 1.4.2):}
    Utilized for various machine learning utilities, including model evaluation metrics.
    \item\textbf{scipy (version 1.15.2):}
     Provided advanced scientific computing functionalities, such as sparse matrix operations and statistical tools.
     \item\textbf{tensorboardX (version 2.6.2.2):}
     Integrated for experiment tracking, visualization of training progress, and logging.
      \item\textbf{torch\_geometric (version 2.6.1): }
    Specifically used for implementing graph neural network components and handling graph data structures.
    \end{itemize} 
\end{itemize}

\subsection{Prediction Performance}
\begin{itemize}
  \item\textbf{Australia-COVID Dataset Prediction Analysis (daily)}

    \begin{table}[h!] 
    \centering
    \scalebox{0.6}{
    \begin{tabular}{llrrrrrrrrrrr} 
        \toprule
        \multirow{2}{*}{\textbf{Method}} & \multirow{2}{*}{\textbf{Metric}} & \multicolumn{11}{c}{\textbf{horizon}} \\
        \cmidrule(lr){3-13} 
        & & \textbf{2} & \textbf{5} & \textbf{8} & \textbf{11} & \textbf{14} & \textbf{17} & \textbf{20} & \textbf{23} & \textbf{26} & \textbf{29} & \textbf{32} \\
        \midrule
        EpiGNN & MAE & 127.2462 & 124.475 & \textbf{100.546} & \textbf{112.6527} & 163.1471 & 170.5775 & 194.5076 & 212.0276 & 210.8035 & 281.8333 & 304.4768 \\
        & RMSE & 388.4988 & 341.19 & \textbf{315.349} & \textbf{370.0045} & \textbf{495.6291} & \textbf{513.5239} & 573.451 & 617.045 & 614.6236 & 725.7447 & 828.0177 \\
        & PCC & 0.9942 & 0.9925 & \textbf{0.9919} & \textbf{0.9905} & \textbf{0.9887} & \textbf{0.9864} & \textbf{0.9863} & \textbf{0.984} & 0.981 & \textbf{0.9816} & 0.9834 \\
        \midrule
        ColaGNN & MAE & \textbf{38.5352} & \textbf{68.965} & 117.77 & 129.8169 & \textbf{149.3325} & \textbf{154.6378} & \textbf{116.4481} & \textbf{142.1623} & \textbf{145.512} & \textbf{129.1289} & \textbf{108.8203} \\
        & RMSE & \textbf{180.3111} & \textbf{276.612} & 434.282 & 458.9208 & 523.0732 & 546.5959 & \textbf{450.783} & \textbf{513.2499} & \textbf{547.5059} & \textbf{499.4233} & \textbf{430.2485} \\
        & PCC & \textbf{0.9974} & \textbf{0.9948} & 0.9882 & 0.9878 & 0.9828 & 0.9798 & 0.9833 & 0.9798 & 0.9747 & 0.9781 & \textbf{0.9841} \\
        \midrule
        STGCN & MAE & 451.3148 & 229.833 & 295.414 & 335.3027 & 305.1303 & 268.4486 & 679.8415 & 190.1576 & 245.7612 & 268.5856 & 413.9621 \\
        & RMSE & 1052.659 & 601.93 & 796.312 & 911.3586 & 842.0427 & 723.162 & 1837.4347 & 572.1667 & 716.6028 & 683.0385 & 919.0851 \\
        & PCC & 0.8412 & 0.9829 & 0.9622 & 0.9763 & 0.9835 & 0.985 & 0.8897 & 0.9833 & \textbf{0.9818} & 0.9743 & 0.8863 \\
        \bottomrule
    \end{tabular}
    }
    \caption{Performance Comparison of Epidemic Prediction Models in Australia. Bold face indicates the best result of each column.}
    \label{tab:australia_performance_bolded}
    \end{table}
    
    \begin{figure}[H]
        \centering
        \begin{minipage}[b]{0.32\textwidth} 
            \includegraphics[width=\linewidth]{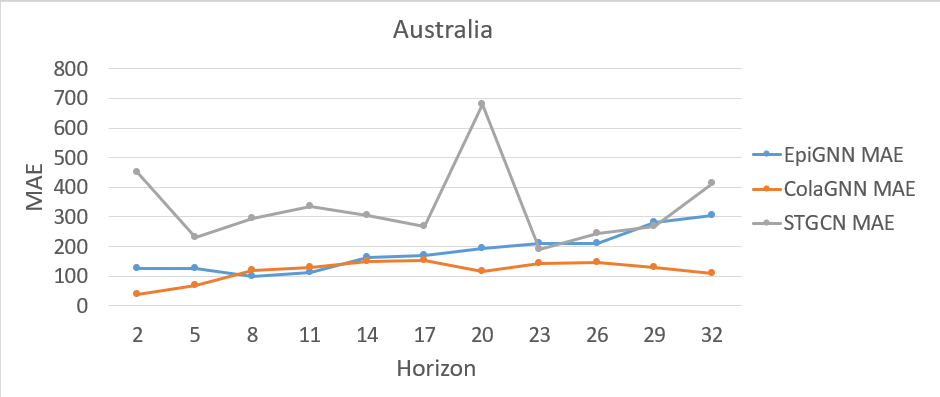}
        \end{minipage}
        \hfill
        \begin{minipage}[b]{0.32\textwidth} 
            \includegraphics[width=\linewidth]{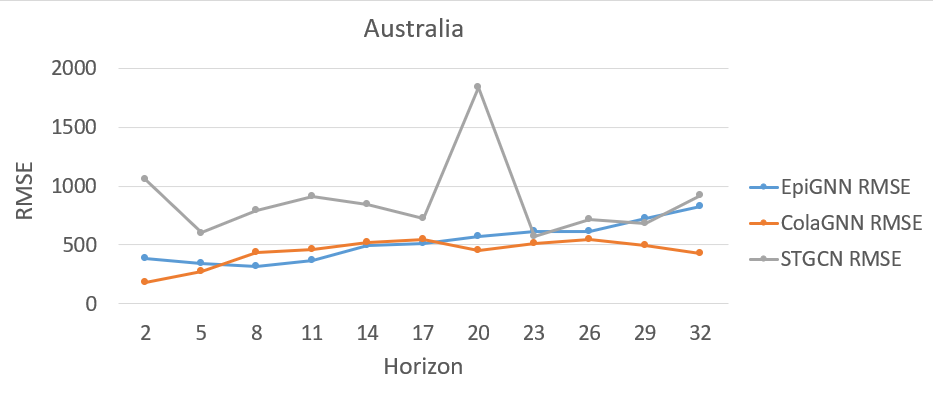}
        \end{minipage}
        \hfill
        \begin{minipage}[b]{0.32\textwidth} 
            \includegraphics[width=\linewidth]{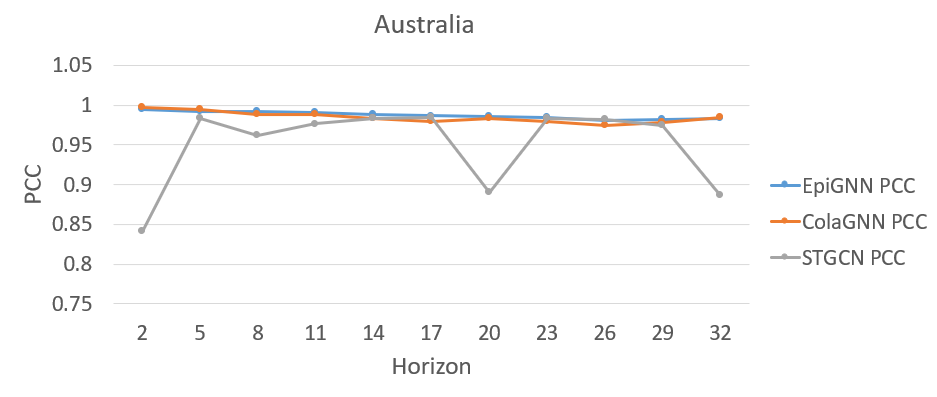}
        \end{minipage}
        \caption{Prediction performance on Australia-COVID dataset}
        \label{fig: Three metrics on Australia-COVID dataset}
    \end{figure}

The analysis of the Australia-COVID dataset reveals distinct strengths and weaknesses across the EpiGNN, ColaGNN, and STGCN models, as detailed in Table \ref{tab:australia_performance_bolded} and further illustrated in Figure \ref{fig: Three metrics on Australia-COVID dataset}. EpiGNN shows its strength in the short-to-mid-term (Horizon 8–17), achieving its lowest error at Horizon 8 and demonstrating a highly consistent PCC that indicates stable trend-fitting. Conversely, ColaGNN is superior for long-term (Horizon > 20) predictions, where it maintains significantly lower errors. Although EpiGNN is strong in a specific mid-range, its errors escalate substantially in later horizons, highlighting ColaGNN's more robust predictive stability across the full forecast window. In contrast, STGCN exhibited drastic fluctuations across all metrics, with large and unstable errors and poor trend correlation, lacking reliable predictive performance.

\item\textbf{Japan Dataset Prediction Analysis (weekly)}

    \begin{table}[h!] 
    \centering
    \scalebox{0.55}{
    \begin{tabular}{llrrrrrrrrrrr} 
        \toprule
        \multirow{2}{*}{\textbf{Method}} & \multirow{2}{*}{\textbf{Metric}} & \multicolumn{11}{c}{\textbf{horizon}} \\
        \cmidrule(lr){3-13} 
        & & \textbf{2} & \textbf{5} & \textbf{8} & \textbf{11} & \textbf{14} & \textbf{17} & \textbf{20} & \textbf{23} & \textbf{26} & \textbf{29} & \textbf{32} \\
        \midrule
        EpiGNN & MAE & 317.7547 & 421.4113 & 534.7141 & 667.254 & 532.3716 & 600.1184 & 740.9535 & 589.0948 & 527.1279 & 572.4716 & 566.4541 \\
        & RMSE & \textbf{934.0154} & \textbf{1051.202} & \textbf{1259.9516} & 1639.9694 & 1464.4508 & 1660.7635 & 1678.3348 & 1403.6164 & 1455.9536 & 1551.2411 & \textbf{1454.3491} \\
        & PCC & 0.9168 & \textbf{0.9016} & 0.8531 & 0.6671 & 0.7321 & 0.6517 & 0.5853 & 0.7966 & 0.792 & 0.722 & 0.7077 \\
        \midrule
        ColaGNN & MAE & \textbf{287.5924} & \textbf{379.4594} & \textbf{412.5771} & \textbf{500.2072} & \textbf{491.446} & \textbf{548.1924} & \textbf{519.0634} & \textbf{499.7023} & \textbf{462.7714} & \textbf{492.9672} & 616.65 \\
        & RMSE & 948.7436 & 1152.2875 & 1275.8528 & \textbf{1421.0287} & \textbf{1412.8741} & \textbf{1484.6521} & \textbf{1471.9136} & \textbf{1388.5836} & 1361.7301 & 1540.4139 & 1657.0711 \\
        & PCC & \textbf{0.9253} & 0.8628 & \textbf{0.8681} & \textbf{0.7744} & \textbf{0.7812} & 0.7339 & 0.8148 & \textbf{0.8098} & \textbf{0.8673} & 0.6967 & 0.595 \\
        \midrule
        STGCN & MAE & 304.3613 & 433.1643 & 449.0197 & 512.6342 & 538.246 & \textbf{534.1789} & 521.1423 & 568.3075 & \textbf{410.4855} & 518.1366 & \textbf{522.0048} \\
        & RMSE & 967.6338 & 1301.5097 & 1353.251 & 1492.7615 & 1598.077 & 1651.6565 & 1565.2718 & 1434.4414 & \textbf{1275.8279} & \textbf{1455.7326} & 1498.1865 \\
        & PCC & 0.9071 & 0.8382 & 0.8306 & 0.7687 & 0.7752 & \textbf{0.7601} & \textbf{0.8233} & 0.7725 & 0.8558 & 0.832 & \textbf{0.757} \\
        \bottomrule
    \end{tabular}
    }
    \caption{Performance Comparison of Epidemic Prediction Models in Japan}
    \label{tab:Japan_performance_bolded}
    \end{table}

    \begin{figure}[H]
        \centering
        \begin{minipage}[b]{0.32\textwidth} 
            \includegraphics[width=\linewidth]{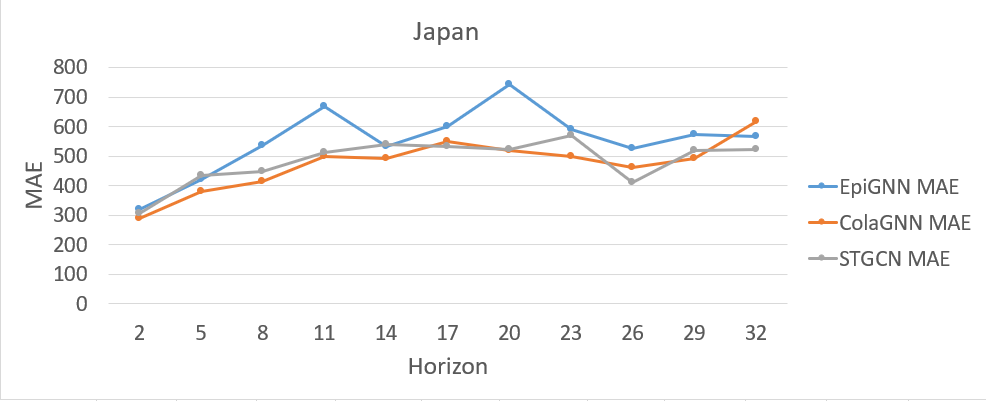}
        \end{minipage}
        \hfill
        \begin{minipage}[b]{0.32\textwidth} 
            \includegraphics[width=\linewidth]{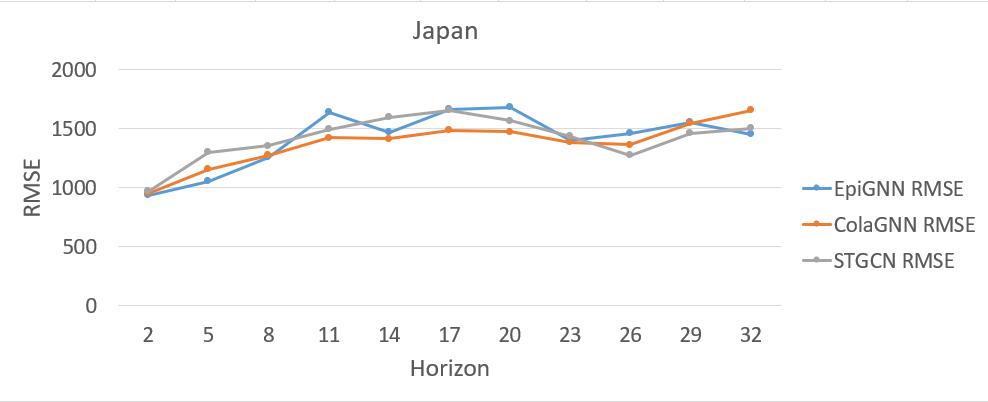}
        \end{minipage}
        \hfill
        \begin{minipage}[b]{0.32\textwidth} 
            \includegraphics[width=\linewidth]{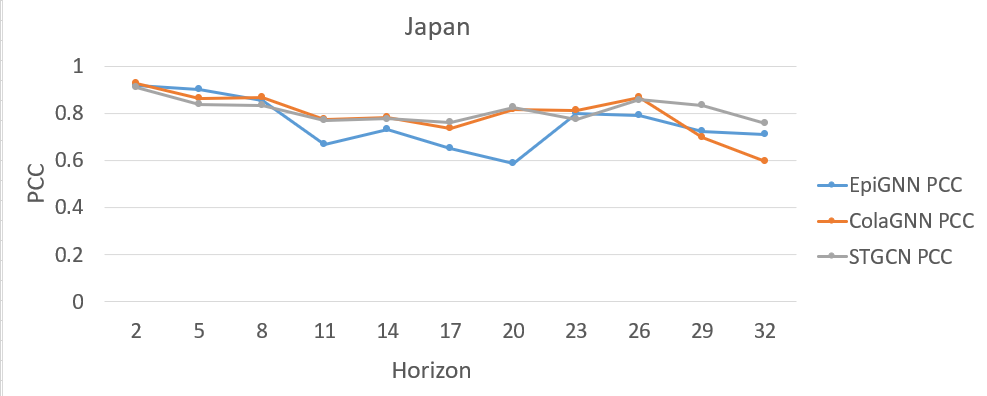}
        \end{minipage}
        \caption{Prediction performance on Japan dataset}
        \label{fig: Three metrics on Japan-Prefectures dataset}
    \end{figure}

The analysis of the Japan-Prefectures dataset is detailed in Table \ref{tab:Japan_performance_bolded} and further illustrated in Figure \ref{fig: Three metrics on Japan-Prefectures dataset}. EpiGNN demonstrates a clear advantage in short-term forecasting (horizons 2, 5, 8), leading in RMSE with values like 934.0154 (horizon 2) and 1051.202 (horizon 5), indicating its ability to minimize large errors in near-term predictions, making it suitable for scenarios requiring precise short-term epidemic trends. ColaGNN consistently demonstrates a strong advantage in MAE across most horizons (especially short to mid-term) and often excels in RMSE and PCC in the mid-term range, showcasing its superior accuracy and correlation for extended horizons. STGCN, while remains competitive and even achieves the lowest RMSE of all models at horizon 26 and 29, is still not as competitive as ColaGNN.
\item\textbf{US-Regions Dataset Prediction Analysis (weekly)}

\begin{table}[H] 
    \centering
    \scalebox{0.55}{
    \begin{tabular}{llrrrrrrrrrrr} 
        \toprule
        \multirow{2}{*}{\textbf{Method}} & \multirow{2}{*}{\textbf{Metric}} & \multicolumn{11}{c}{\textbf{horizon}} \\
        \cmidrule(lr){3-13} 
        & & \textbf{2} & \textbf{5} & \textbf{8} & \textbf{11} & \textbf{14} & \textbf{17} & \textbf{20} & \textbf{23} & \textbf{26} & \textbf{29} & \textbf{32} \\
        \midrule
        EpiGNN & MAE & 271.9734 & 543.7277 & \textbf{608.6584} & \textbf{693.9379} & \textbf{584.1873} & \textbf{595.4306} & 812.4983 & 743.6681 & 815.2423 & 679.3557 & 689.5122 \\
        & RMSE & \textbf{495.2303} & \textbf{898.0375} & \textbf{1024.3151} & \textbf{1139.753} & \textbf{994.9219} & \textbf{990.6613} & 1278.9443 & 1203.0116 & 1292.1929 & \textbf{1043.0071} & 989.5857 \\
        & PCC & 0.9421 & 0.7813 & \textbf{0.7064} & \textbf{0.6038} & \textbf{0.7312} & \textbf{0.7339} & 0.5564 & 0.5585 & 0.5114 & 0.6884 & 0.7704 \\
        \midrule
        ColaGNN & MAE & \textbf{233.6242} & \textbf{525.8035} & 685.4155 & 786.2728 & 708.7531 & 848.0165 & 878.642 & \textbf{658.7186} & \textbf{633.9754} & \textbf{484.4075} & \textbf{447.5002} \\
        & RMSE & 506.9187 & 1007.5835 & 1270.2356 & 1407.126 & 1260.4982 & 1429.4681 & 1483.2721 & \textbf{1061.4983} & \textbf{1206.7821} & \textbf{969.8692} & \textbf{904.5961} \\
        & PCC & \textbf{0.9435} & \textbf{0.7988} & 0.6637 & 0.6036 & 0.6874 & 0.46159 & 0.4159 & \textbf{0.6701} & \textbf{0.7355} & \textbf{0.774} & \textbf{0.7968} \\
        \midrule
        STGCN & MAE & 408.5528 & 556.1506 & 701.1787 & 784.0956 & 800.3231 & 790.7985 & \textbf{684.8737} & 768.8422 & 684.4204 & 587.5015 & 523.9285 \\
        & RMSE & 697.2763 & 1066.6802 & 1267.289 & 1384.2491 & 1381.3204 & 1413.0061 & \textbf{1235.4878} & 1305.1195 & 1242.937 & 1077.7383 & 996.238 \\
        & PCC & 0.8783 & 0.7682 & 0.6872 & 0.5907 & 0.4352 & 0.6718 & \textbf{0.7169} & 0.5499 & 0.7164 & 0.7574 & 0.7571 \\
        \bottomrule
    \end{tabular}
    }
    \caption{Performance Comparison of Epidemic Prediction Models in US-Regions} 
    \label{tab:performance_comparison_v2} 
\end{table}

\begin{figure}[H]
        \centering
        \begin{minipage}[b]{0.32\textwidth} 
            \includegraphics[width=\linewidth]{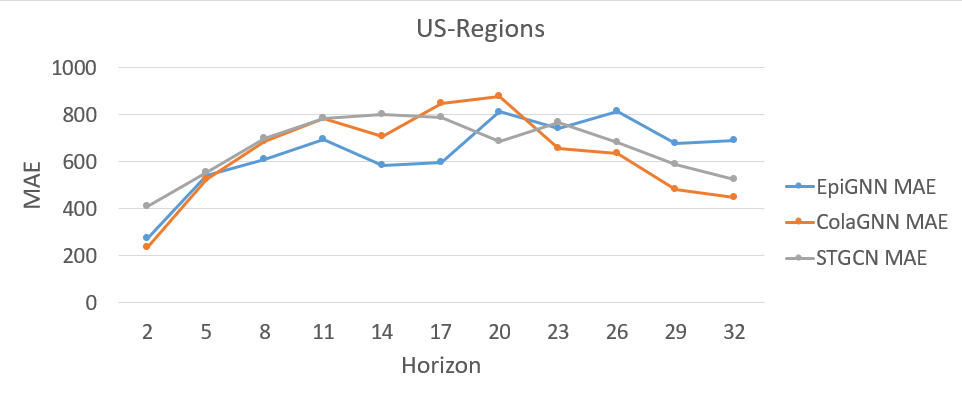}
        \end{minipage}
        \hfill
        \begin{minipage}[b]{0.32\textwidth} 
            \includegraphics[width=\linewidth]{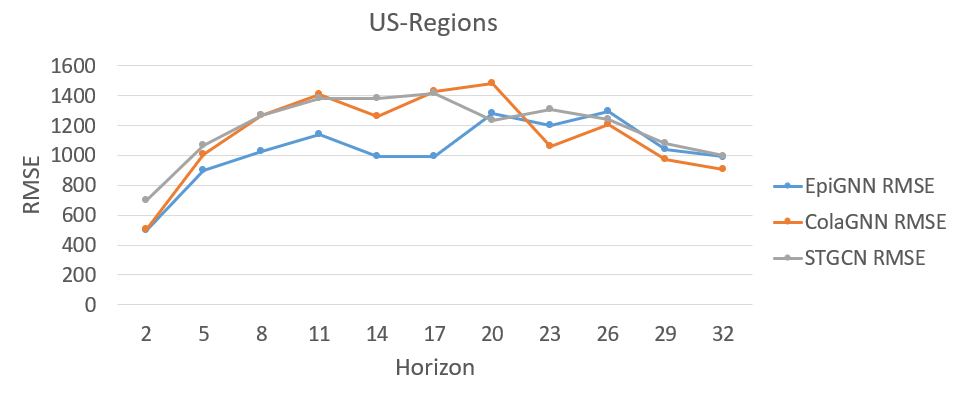}
        \end{minipage}
        \hfill
        \begin{minipage}[b]{0.32\textwidth} 
            \includegraphics[width=\linewidth]{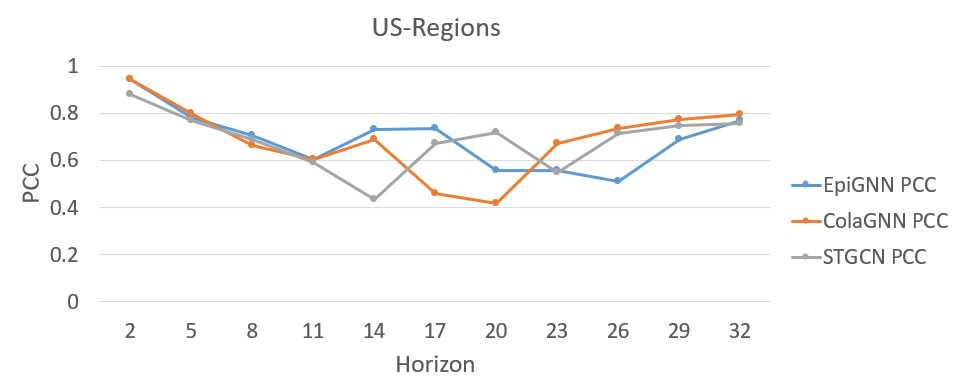}
        \end{minipage}
        \caption{Prediction performance on US-Regions dataset}
        \label{fig:Three metrics on US-Regions}
    \end{figure}

The analysis of the US-Regions dataset is detailed in Table \ref{tab:performance_comparison_v2} and further illustrated in Figure \ref{fig:Three metrics on US-Regions}. EpiGNN demonstrates a notable advantage in the short-term prediction phase (Horizon $\leq$ 17). In this range, its RMSE is consistently lower than the other models, such as 990.6613 at horizon 17 compared to ColaGNN's 1429.4681, indicating stronger predictive accuracy. However, as the time horizon extends, ColaGNN establishes clear superiority in the medium to long-term (Horizon > 20). In this later phase, its MAE and RMSE are significantly lower and more stable than EpiGNN's and STGCN's. For instance, at horizon 32, ColaGNN’s RMSE is 904.5961, while EpiGNN’s is 989.5857. Furthermore, its PCC shows a steady improvement in the long term, suggesting better trend consistency. In contrast, STGCN remains volatile across all metrics, exhibiting high errors and multiple correlation troughs, which indicates weak and unreliable predictive performance on this dataset.

\item\textbf{US-States Dataset Prediction Analysis (weekly)}

\begin{table}[h!] 
    \centering
    \scalebox{0.55}{
    \begin{tabular}{llrrrrrrrrrrr} 
        \toprule
        \multirow{2}{*}{\textbf{Method}} & \multirow{2}{*}{\textbf{Metric}} & \multicolumn{11}{c}{\textbf{horizon}} \\
        \cmidrule(lr){3-13} 
        & & \textbf{2} & \textbf{5} & \textbf{8} & \textbf{11} & \textbf{14} & \textbf{17} & \textbf{20} & \textbf{23} & \textbf{26} & \textbf{29} & \textbf{32} \\
        \midrule
        EpiGNN & MAE & 58.6324 & 85.646 & 95.9525 & 94.9318 & 107.6102 & 105.6782 & 112.834 & 108.9413 & 107.7824 & 138.4842 & 141.8843 \\
        & RMSE & 148.2381 & \textbf{199.371} & \textbf{212.59} & \textbf{210.8205} & \textbf{228.3557} & \textbf{231.091} & \textbf{234.3161} & \textbf{236.3079} & 227.6159 & 265.86 & 281.6397 \\
        & PCC & 0.9449 & \textbf{0.8975} & \textbf{0.8827} & \textbf{0.8868} & 0.8601 & 0.8602 & \textbf{0.8515} & 0.8508 & 0.8607 & 0.8088 & 0.7776 \\
        \midrule
        ColaGNN & MAE & \textbf{53.4726} & \textbf{83.3606} & \textbf{91.9288} & \textbf{94.4864} & \textbf{95.2117} & \textbf{97.5189} & \textbf{108.0375} & \textbf{101.7764} & \textbf{97.9957} & \textbf{102.4193} & \textbf{101.776} \\
        & RMSE & \textbf{148.1823} & 221.7605 & 223.09 & 227.3539 & 229.4957 & 232.3823 & 236.4327 & 238.5425 & \textbf{224.7046} & \textbf{233.7741} & \textbf{241.2217} \\
        & PCC & \textbf{0.9498} & 0.8747 & 0.8723 & 0.8671 & \textbf{0.8797} & \textbf{0.8816} & 0.849 & \textbf{0.8663} & \textbf{0.8744} & \textbf{0.8613} & \textbf{0.8545} \\
        \midrule
        STGCN & MAE & 82.7078 & 101.3556 & 111.7042 & 113.3681 & 112.4669 & 111.6197 & 110.8818 & 11.9671 & 112.5974 & 108.3811 & 107.0486 \\
        & RMSE & 181.8974 & 235.4713 & 270.4545 & 267.4492 & 271.1715 & 266.1554 & 269.5841 & 254.2331 & 255.5211 & 244.7531 & 245.1524 \\
        & PCC & 0.914 & 0.8517 & 0.8036 & 0.807 & 0.8092 & 0.8115 & 0.8138 & 0.8281 & 0.8255 & 0.8453 & 0.8513 \\
        \bottomrule
    \end{tabular}
    }
    \caption{Performance Comparison of Epidemic Prediction Models Across in US-States} 
    \label{tab:performance_comparison_v3} 
\end{table}

\begin{figure}[H]
        \centering
        \begin{minipage}[b]{0.32\textwidth} 
            \includegraphics[width=\linewidth]{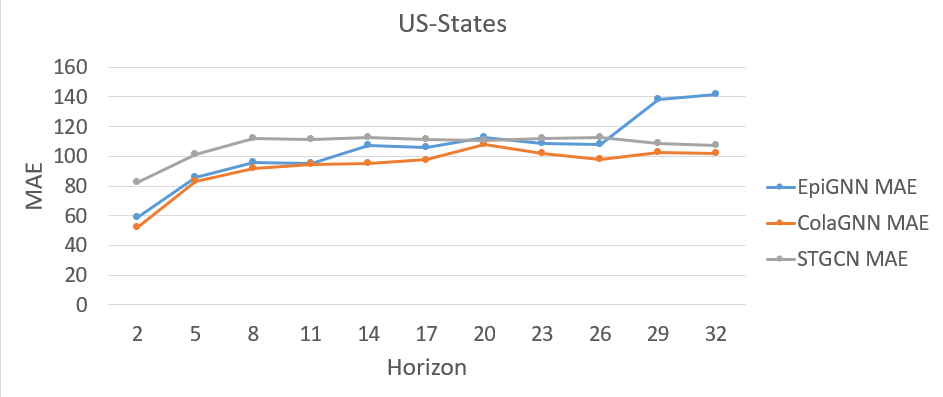}
        \end{minipage}
        \hfill
        \begin{minipage}[b]{0.32\textwidth} 
            \includegraphics[width=\linewidth]{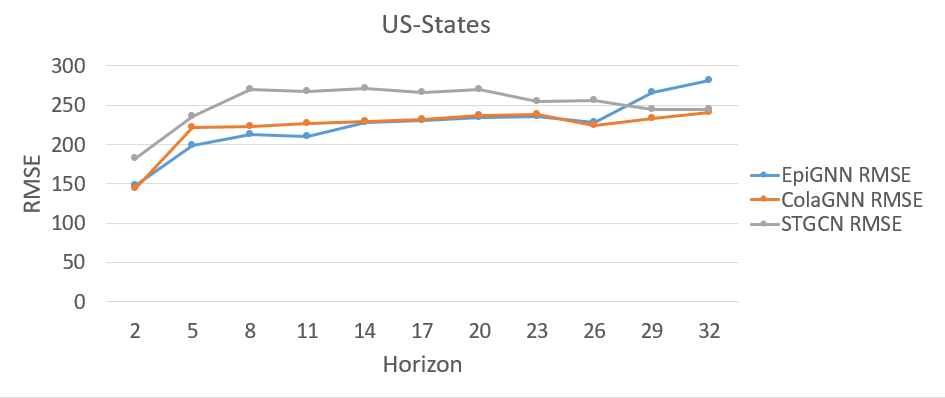}
        \end{minipage}
        \hfill
        \begin{minipage}[b]{0.32\textwidth} 
            \includegraphics[width=\linewidth]{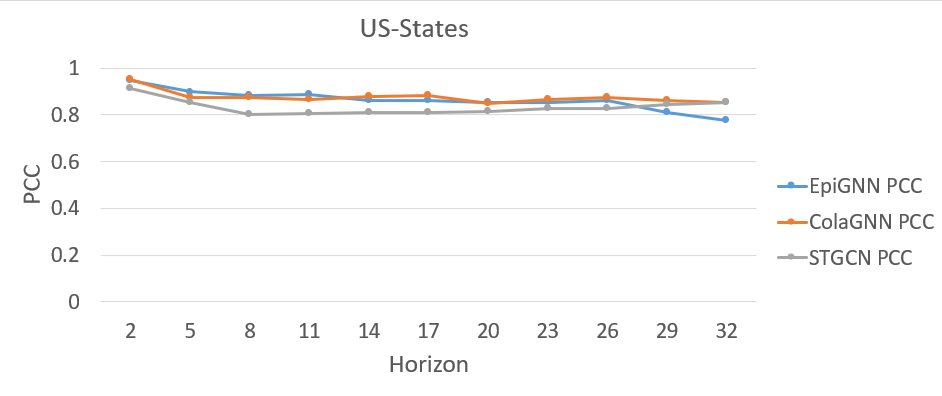}
        \end{minipage}
        \caption{Prediction performance on US-States dataset}
        \label{fig:Three metrics on US-States}
    \end{figure}
The analysis of the US-States dataset is detailed in Table \ref{tab:performance_comparison_v3} and further illustrated in Figure \ref{fig:Three metrics on US-States}. ColaGNN and EpiGNN exhibit highly competitive performance, especially in the short term. ColaGNN consistently achieves the lowest MAE across nearly all stages and demonstrates superior performance in the long term (Horizon $\geq$ 26), where its RMSE remains lower and more stable, and its PCC is consistently higher. This makes it well-suited for stable short-to-long term trend modeling. EpiGNN is very competitive, even outperforming ColaGNN in RMSE during the short-term phase (Horizon 5-11), but its error increases more significantly in the long term (Horizon $\geq$ 29), making it most suitable for short-term, high-precision scenarios. In contrast, STGCN's performance is consistently weaker, with the highest errors and lowest correlation across most of the prediction horizon, indicating insufficient overall predictive stability for this dataset.

\textbf{Summary:}
Overall, EpiGNN and ColaGNN each have their strengths, while STGCN performed poorly in this evaluation:
    \begin{itemize}
        \item\textbf{EpiGNN:} Excels in short-term forecasting, showing high accuracy in specific forecast windows on the Australia-COVID, Japan-Prefectures, US-Regions and US-States datasets. Its primary strength lies in high-precision, short-range predictions. However, its performance tends to degrade as the forecast horizon extends, with increasing errors in the long term across most datasets.
        \item\textbf{ColaGNN:} Demonstrates stronger and more consistent overall performance, establishing itself as the more robust model, particularly for medium to long-term predictions. It achieved the most stable and accurate results in the later stages across all datasets. This makes ColaGNN the preferable choice for tasks requiring stable medium-to-long-term forecasting and reliable overall trend modeling.
        \item\textbf{STGCN:} Exhibited drastic fluctuations, large errors, and instability in most tests, with poor trend correlation, lacking reliable predictive performance.
    \end{itemize}
Therefore, EpiGNN is preferred for short-term high-precision and handling complex propagation patterns, while ColaGNN performs better in long-term stable forecasting and overall trend modeling.
\end{itemize}

\section{Advantages and Disadvantages Comparison}
\subsection{Advantages and Disadvantages of EpiGNN}

\textbf{Advantages:}
\begin{itemize}
    \item \textbf{Strong spatio-temporal modeling capability:} Effectively captures temporal and spatial dependencies through multi-scale convolutions and GCNs. The RAGL module further integrates geographical information and propagation risk, generating an asymmetric correlation graph that avoids the over-smoothing problem of traditional attention mechanisms, performing exceptionally well in complex propagation paths.
    \item \textbf{Combination of linear and non-linear components:} By combining AR components and GCN outputs, it balances linear patterns (e.g., government intervention effects in COVID-19) and non-linear patterns (e.g., seasonal fluctuations in influenza), enhancing the model's robustness.
    \item \textbf{Flexible external resource integration:} The model supports integrating external data such as human mobility, improving the accuracy of inter-regional correlations.
    \item \textbf{Interpretability and risk identification:} LTR and GTR modules provide interpretability of regional propagation risks through quantified indicators, aiding in identifying high-risk areas.
\end{itemize}
\textbf{Disadvantages:}
\begin{itemize}
    \item \textbf{Decreased long-term prediction performance:} Experiments show that as the prediction horizon $h$ increases, prediction accuracy decreases, indicating an imbalance in multi-source dynamic inputs over long sequences and limited ability to capture long-term propagation dynamics.
    \item \textbf{Complex parameter tuning:} Multi-scale convolutions, GCN layer counts, and RAGL hyperparameters require careful tuning, increasing model deployment complexity.
    \item \textbf{Lack of temporal decay effect:} The model does not account for the characteristic that propagation influence decays over time, which may affect the accuracy of long-term predictions.
\end{itemize}

\subsection{Advantages and Disadvantages of ColaGNN}
\textbf{Advantages:}
\begin{itemize}
    \item \textbf{Long-term forecasting capability:} Relies on RNN recursive modeling and a location-aware attention matrix, suitable for medium to long-term trend modeling.
    \item \textbf{Dynamic graph structure:} The cross-location attention mechanism can dynamically capture complex dependencies between non-adjacent regions.
    \item \textbf{Strong interpretability:} The attention matrix helps epidemiologists understand disease diffusion patterns.
    \item \textbf{High computational efficiency:} Small number of parameters and fast training speed, suitable for forecasting tasks involving large-scale regions.
\end{itemize}
\textbf{Disadvantages:}
\begin{itemize}
    \item \textbf{Computational overhead of attention mechanism:} Although the overall number of parameters is relatively small, the $N^2$-level computational overhead of the attention mechanism can still be a challenge when processing a large number of regions.
    \item \textbf{Risk of overfitting:} The attention mechanism relies on the direct computation of the last hidden state of the RNN. In data-sparse or small-sample scenarios (e.g., some regions in US-Regions), the model may overfit local patterns.
    \item \textbf{Lack of explicit risk encoding:} Unlike EpiGNN, it does not explicitly encode local and global transmission risks; this information may need to be implicitly learned from spatio-temporal sequences.
\end{itemize}

\subsection{Model Comparison}
\textbf{Overall Comparison:} Both EpiGNN and ColaGNN employ multi-scale convolutions to capture temporal patterns and leverage GNN structures to handle spatial dependencies. Both emphasize the importance of dynamic graph structures and attention mechanisms in capturing complex propagation patterns. EpiGNN's highlight lies in its transmission risk encoding module and Region-Aware Graph Learner (RAGL), which allows it to more finely integrate local and global propagation risks and geographical information, performing exceptionally well in epidemics like COVID-19 that are significantly influenced by government interventions. Additionally, EpiGNN's design combining linear AR components enhances its robustness. ColaGNN, on the other hand, focuses more on long-term forecasting, and its cross-location attention mechanism can dynamically capture complex dependencies between non-adjacent regions, which is crucial for predicting long-term disease propagation patterns. ColaGNN also performs well in terms of data efficiency and model complexity. Nevertheless, both models have their limitations. EpiGNN is limited in long-term forecasting, and its hyperparameter tuning is complex. ColaGNN, while excellent in long-term forecasting, does not offer advantages in short-term predictions and lacks more direct risk encoding to guide epidemiological research.

\section{Improvement Space and A HybridGNN Model}
\subsection{Improvement Space}
EpiGNN and ColaGNN have significant complementarity in their design philosophies and application focuses:
\begin{itemize}
    \item \textbf{Short-term and long-term prediction complementarity:} EpiGNN excels in short-term forecasting and handling sudden outbreaks, and its ability to integrate external events allows it to better adapt to rapid changes in the early stages of an epidemic. ColaGNN, on the other hand, focuses on long-term forecasting, and its dilated convolutions and dynamic attention mechanism are more suitable for capturing periodic, seasonal, and other long-term patterns.
    \item \textbf{Explicit risk encoding versus dynamic attention:} EpiGNN explicitly designs a transmission risk encoding module, incorporating local and global propagation risks into the model, which provides a more direct perspective for understanding the "causes" and "effects" of disease propagation. While ColaGNN's dynamic attention mechanism can also capture this information, it learns implicitly, making it slightly less interpretable. Combining the two can achieve more comprehensive risk awareness.
    \item \textbf{External data integration complementarity:} EpiGNN explicitly supports the integration of external data, which is highly beneficial for scenarios with abundant external data. ColaGNN currently primarily relies on existing data, but its dynamic graph learning framework can, in principle, also integrate similar information. Combining EpiGNN's external data integration capabilities with ColaGNN's efficient dynamic graph learning can build a more powerful model.
\end{itemize}
Based on the strong complementarity between the two in these three aspects, a new model structure can be designed to integrate their respective advantages and compensate for their shortcomings, thereby achieving more effective epidemic forecasting. The EpiCola-GNN model integrates EpiGNN's strengths in risk encoding and local-aware convolutions with ColaGNN's characteristics in long-term forecasting and dynamic attention mechanisms, forming a unified, adaptive spatio-temporal data prediction model designed to handle complex spatio-temporal propagation problems.

\subsection{EpiCola-GNN Structure Overview}
The core design idea of EpiCola-GNN is:

\begin{itemize}
    \item \textbf{Multi-scale temporal feature extraction:} Utilizes ColaGNN's multi-scale dilated convolutions, which include short-term and long-term receptive fields, to comprehensively capture temporal features from raw time series data, crucial for disease propagation and other long-term prediction tasks.
    \item \textbf{Transmission risk encoding:} Borrows from EpiGNN's Global Transmission Risk (GTR) and Local Transmission Risk (LTR) modules, quantifying and encoding potential propagation risks between regions through attention mechanisms and geographical degree information.
    \item \textbf{Dynamic graph construction:} Combines ColaGNN's RNN-based temporal hidden state-generated location-aware attention mechanism with normalized geographical adjacency matrices and EpiGNN's degree gating mechanism to dynamically construct and adaptively adjust inter-regional connection strengths, forming the final graph structure.
    \item \textbf{Feature fusion and graph message passing:} Concatenates multi-scale temporal features, LTR, and GTR encodings to form initial node features for each region. These features are then propagated and learned through multiple layers of Graph Convolutional Networks (GCNs) on the dynamically constructed graph structure.
    \item \textbf{Multi-layer GCN propagation:} Employs a multi-layer GCN architecture, enabling node features to effectively aggregate neighbor information and transform node features under the guidance of the dynamic Laplacian matrix. The model supports residual connections to preserve information learned at each layer.
    \item \textbf{Integrated prediction:} In the final prediction stage, the spatio-temporal features processed by GCN are fused with the pure temporal hidden states extracted by RNN. Additionally, the model can optionally include residual connections based on historical raw input data to enhance prediction accuracy and stability.
\end{itemize}

\subsection{EpiCola-GNN Module Details}
The specific modules of the model and their operations are shown in Figure \ref{fig:EpiCola-GNN workflow}:
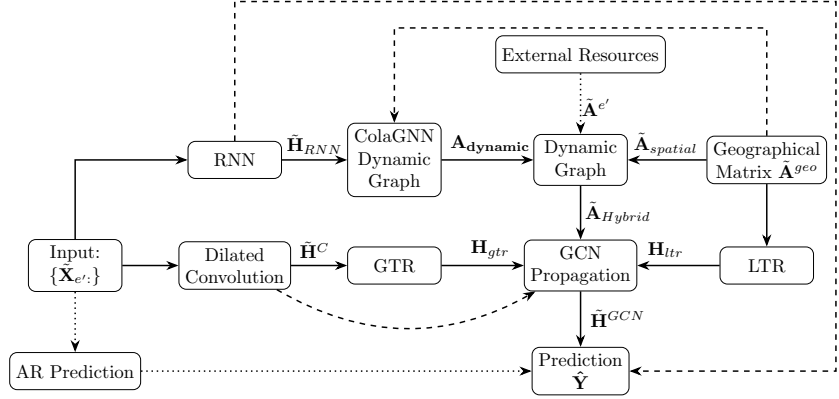
\begin{figure}[H]
        \centering
        \scalebox{0.7}{
        \begin{tikzpicture}[
            box/.style={rectangle, draw, rounded corners, minimum height=2em, minimum width=5em, align=center},
            arrow/.style={-Stealth, thick}
        ]
            \node[box] (input) at (0,0) {Input: \\$\{\tilde{\mathbf{X}}_{e^{\prime}:}\}$};
            \node[box] (dc) at (3,0) {Dilated\\ Convolution};
            \node[box] (gtr) at (6,0) {GTR};
            \node[box] (gcn) at (9.5,0) {GCN\\Propagation};
            \node[box] (ltr) at (13,0) {LTR};
            \node[box] (rnn) at (3,2) {RNN};
            \node[box] (cola) at (6,2) {ColaGNN\\ Dynamic \\Graph};
            \node[box] (dy) at (9.5,2) {Dynamic\\ Graph};
            \node[box] (geo) at (13,2) {Geographical \\Matrix $\tilde{\mathbf{A}}^{geo}$};
            \node[box] (er) at (9.5,4) {External Resources};
            \node[box] (predict) at (9.5,-2) {Prediction\\$\mathbf{\hat{Y}}$};
            \node (hrnn) at (4.5,2.3) {$\tilde{\mathbf{H}}_{RNN}$};
            \node (ady) at (7.8,2.3) {$\mathbf{A_{dynamic}}$};
            \node (aspa) at (11.1,2.3) {$\tilde{\mathbf{A}}_{spatial}$};
            \node (hg) at (7.8,0.3) {$\mathbf{H}_{gtr}$};
            \node (hl) at (11.1,0.3) {$\mathbf{H}_{ltr}$};
            \node (aa) at (10.2,1) {$\tilde{\mathbf{A}}_{Hybrid}$};
            \node (ae) at (9.8,3) {$\tilde{\mathbf{A}}^{e^{\prime}}$};
            \node (hc) at (4.5,0.3) {$\tilde{\mathbf{H}}^C$};
            \node (hgcn) at (10.2,-1) {$\tilde{\mathbf{H}}^{GCN}$};
            \node[box] (arpredict) at (0,-2) {AR Prediction};

            \draw[arrow, dotted] (input) -- (arpredict);
            \draw[arrow, dotted] (arpredict) -- (predict);
            \draw[arrow] (input) -- (dc);
            \draw[arrow] (input) |- (rnn);
            \draw[arrow] (dc) -- (gtr);
            \draw[arrow] (gtr) -- (gcn);
            \draw[arrow] (ltr) -- (gcn);
            \draw[arrow] (gcn) -- (predict);
            \draw[arrow] (rnn) -- (cola);
            \draw[arrow] (cola) -- (dy);
            \draw[arrow] (geo) -- (dy);
            \draw[arrow, dotted] (er) -- (dy);
            \draw[arrow] (dy) -- (gcn);
            \draw[arrow] (geo) -- (ltr);
            \draw[arrow, dashed] (dc) to[bend right] (gcn);
            \draw[arrow, dashed] (geo) |- ++(0,2.5) -| (cola); 
            \draw[arrow, dashed] (rnn) |- ++(11.3,3) |- (predict); 

        \end{tikzpicture}
        }
        \caption{EpiCola-GNN workflow}
        \label{fig:EpiCola-GNN workflow}
\end{figure}

\begin{itemize}
    \item \textbf{Convolutional Temporal Feature Extraction (based on Cola-GNN)}\\
    This module extracts short-term and long-term temporal patterns from the time series $\hat{\mathbf{X}} \in \mathbb{R}^{B \times T \times N}$ for $N$ regions, batch size $B$, and time window $T$, using one-dimensional dilated convolutions. For each region $e' = 1, \ldots, N$, the time series $\hat{\mathbf{X}}_{e':} \in \mathbb{R}^{B \times T \times 1}$ is reshaped to $\tilde{\mathbf{X}}_{e':} \in \mathbb{R}^{B \times 1 \times T}$ for convolution:
    \begin{equation}
        \tilde{\mathbf{h}}_{e^{\prime}}^{C} = \text{Flatten}(\text{Concat}(\text{Conv1d}_{short}(\tilde{\mathbf{X}}_{e^{\prime}:}), \text{Conv1d}_{long}(\tilde{\mathbf{X}}_{e^{\prime}:})))
    \end{equation}
    Here, $\text{Conv1d}_{short}$ uses $k^{\prime\prime}$ filters with kernel size $T$ and dilation $r^{\prime\prime}
    = 1$, producing an output of shape $\mathbb{R}^{B \times k^{\prime\prime} \times 1}$. $\text{Conv1d}_{long}$ uses $k^{\prime\prime}$ filters with kernel size $\lfloor T/2 \rfloor$ and dilation $r^{\prime\prime} = 2$, producing an output of shape $\mathbb{R}^{B \times k^{\prime\prime}\times(T - 2(\lfloor T/2 \rfloor - 1))}$. Concatenation and flattening yield $\tilde{\mathbf{h}}_{e'}^{C} \in \mathbb{R}^{B \times (k^{\prime\prime}\cdot(1 + (T - 2(\lfloor T/2 \rfloor - 1))))}$. Then, the temporal features of all regions are stacked along the region dimension:
    \begin{equation}        
        \tilde{\mathbf{H}}^C = \text{ReLU}(\text{Stack}([\tilde{\mathbf{h}}_{1}^{C}, \ldots, \tilde{\mathbf{h}}_{N}^{C}])) \in \mathbb{R}^{B \times N \times D_c}
    \end{equation}
    where $D_c = k^{\prime\prime}(1 + (T - 2(\lfloor T/2 \rfloor - 1)))$. 
    \item \textbf{Dynamic Temporal Feature Extraction (RNN)}\\
    For dynamic temporal features, $\hat{\mathbf{X}}$ is reshaped to $\mathbf{X}_{\text{reshape}} \in \mathbb{R}^{(B \cdot N) \times T \times 1}$ and fed to a RNN (LSTM, GRU, or RNN):
    \begin{equation}
        \tilde{\mathbf{H}}_{RNN} = \text{Reshape}\left(\text{RNN}(\mathbf{X}_{\text{reshape}})[:, -1, :]\right)
    \end{equation}
    The RNN, with hidden size $n_{hidden}$, and bidirectional flag $bi$, outputs $\mathbb{R}^{(B \cdot N) \times T \times D_h}$, where $D_h = n_{hidden} \cdot (1 + bi)$. The last hidden state is abstracted by $[:, -1, :]$ and then reshaped to $\tilde{\mathbf{H}}_{\text{RNN}} \in \mathbb{R}^{B \times N \times D_h}$, used for dynamic graph construction and final prediction.
    
    \item \textbf{Transmission Risk Encoding (based on EpiGNN)}\\
    Given the region-wise degree vector $\mathbf{d} \in \mathbb{R}^{N}$ derived from the geographical adjacency matrix $\mathbf{A}^{geo}$, and the temporal features $\tilde{\mathbf{H}}^C$, we calculate both the local transmission risk $\mathbf{H}_{ltr}$ and global transmission risk $\mathbf{H}_{gtr}$.
    \begin{equation}
        \mathbf{H}_{ltr} = \text{Linear}(\text{Expand}(\mathbf{d}))\in \mathbb{R}^{B \times N \times {hidR}} \quad 
    \end{equation}  
    where $hidR$ is the feature embedding dimension for transmission risk encoding. $\text{Expand}(\mathbf{d}) \in \mathbb{R}^{B \times N \times 1}$ broadcasts $\mathbf{d}$ across batch size $B$ to make it suitable for batch processing and $\text{Linear}(\cdot)$ is a linear transformation mapping $\mathbb{R}^{1} \to \mathbb{R}^{hidR}$, yielding $\mathbf{H}_{ltr} \in \mathbb{R}^{B \times N \times {hidR}}$.
    \begin{equation}
       \mathbf{Attn_{gtr}} = \text{Normalize}((\tilde{\mathbf{H}}^C \mathbf{W_Q})(\tilde{\mathbf{H}}^C \mathbf{W_K})^T, \text{dim=-1})
    \end{equation} 
    \begin{equation}
       \mathbf{H}_{gtr} = \text{Linear}(\text{Sum}(\mathbf{Attn_{gtr}}, \text{dim=-1}))\in \mathbb{R}^{B \times N \times {hidR}}.
    \end{equation} 
    
    For global risk, $\mathbf{W_Q}, \mathbf{W_K} \in \mathbb{R}^{D_c \times {hidA}}$ are weight matrices projecting $\tilde{\mathbf{H}}^C$ to $\mathbb{R}^{B \times N \times {hidA}}$, producing query and key matrices. $ hidA $ is the attention hidden dimension for global transmission risk. Their batch matrix product yields $\mathbf{Attn_{gtr}} \in \mathbb{R}^{B \times N \times N}$. $\text{Normalize($\cdot$, dim=-1)}$ performs $L_2$ normalization along the last dimension. $\text{Sum($\cdot$,dim=-1)}$ summing along the last dimension gives $\mathbb{R}^{B \times N \times 1}$, which $\text{Linear}(\cdot)$ maps to $\mathbf{H}_{gtr} \in \mathbb{R}^{B \times N \times {hidR}}$.
    
    \item\textbf{Dynamic Graph Construction (fusion of EpiGNN and ColaGNN)}\\
    This module uses ColaGNN's RNN and additive attention mechanisms fused with EpiGNN geographical adjacency and the external relation matrix to generate a dynamic adjacency matrix for GNN propagation, combining temporal dynamics and geographical priors.

    \begin{itemize}
        \item\textbf{ColaGNN-style Dynamic Graph ($\mathbf{A_{dynamic}}$):}\\
    Using $\tilde{\mathbf{H}}_{RNN} \in \mathbb{R}^{B \times N \times D_h}$ from the temporal module, the additive attention is:
    \begin{equation}
       \mathbf{A}^P_{ij} = \text{Normalize}(\text{ELU}( (\tilde{\mathbf{H}}_{RNN})_i \mathbf{W^T_{p_1}} + (\tilde{\mathbf{H}}_{RNN})_j \mathbf{W^T_{p_2}} + \mathbf{b_{p_1}} ) \mathbf{V_p} + b_{pv}) \in \mathbb{R}^{B \times 1}
    \end{equation} 
    Here,  $(\tilde{\mathbf{H}}_{RNN})_i \in \mathbb{R}^{B \times 1 \times D_h}$ is the hidden state for location $i$. $\mathbf{W^{T}_{p_1}}, \mathbf{W^{T}_{p_2}} \in \mathbb{R}^{D_h \times \text{half}\_{hid}}$, $\mathbf{b_{p1}} \in \mathbb{R}^{\text{half}\_{hid}}$, $\mathbf{V_p} \in \mathbb{R}^{\text{half}\_{hid}}$, $\text{half}\_{hid}= \left\lfloor \frac{\text{n}_{hidden}}{2} \right\rfloor$ and $b_{pv} \in \mathbb{R}^1$ are trainable parameters. The attention scores form a batched matrix $\mathbf{A}^P \in \mathbb{R}^{B \times N \times N}$ that undergoes $L_2$ normalization. The batched gating matrix is:
    \begin{equation}
       \mathbf{G}^{*} = \text{Sigmoid}(\mathbf{A}^P \mathbf{W_b} + w_b) 
    \end{equation}
    where $\mathbf{W_b} \in \mathbb{R}^{N \times N}$, $w_b \in \mathbb{R}$, are trainable parameters yielding $\mathbf{G}^{*} \in \mathbb{R}^{B \times N \times N}$. The dynamic adjacency is:
    \begin{equation}
       \mathbf{A_{dynamic}} = (\mathbf{A}^G \odot \mathbf{G}^{*}) + (\mathbf{A}^P \odot (1-\mathbf{G}^{*}))\in \mathbb{R}^{B \times N \times N}
    \end{equation} 
    where $\mathbf{A}^G = \text{Expand}(\tilde{\mathbf{A}}^{geo})\in \mathbb{R}^{B \times N \times N}$ is the batched normalized geographical adjacency matrix.

    \item\textbf{EpiGNN-style Spatial Adjacency:}\\
    The degree vector $\mathbf{d} \in \mathbb{R}^{N \times 1}$ is expanded to $\text{Expand}(\mathbf{d}) \in \mathbb{R}^{B \times N \times 1}$. The gating term is:
    \begin{equation}
       \mathbf{D_{mat}} = \text{Sigmoid}(\mathbf{W}^{gate} \odot (\text{Expand}(\mathbf{d}) \text{Expand}(\mathbf{d})^T))
    \end{equation} 
    where $\mathbf{W}^{gate} \in \mathbb{R}^{N \times N}$, $\text{Expand}(\mathbf{d}) \text{Expand}(\mathbf{d})^T \in \mathbb{R}^{B \times N \times N}$, yielding $\mathbf{D_{mat}} \in \mathbb{R}^{B \times N \times N}$. The spatial adjacency is obtained by:
    \begin{equation}
       \tilde{\mathbf{A}}_{spatial} = \mathbf{D_{mat}} \odot \mathbf{A}^G \in \mathbb{R}^{B \times N \times N}
    \end{equation}

    \item\textbf{Final Fusion:}
    \begin{equation}\label{eq:epicola}
       \tilde{\mathbf{A}}_{Hybrid} = \mathbf{A_{dynamic}} + \tilde{\mathbf{A}}_{spatial} + \tilde{\mathbf{A}}^{e^{\prime}}
    \end{equation} 

    Here, $\tilde{\mathbf{A}}^{e^{\prime}} =\text{Expand}(\mathbf{A}^{e^{\prime}}) \in \mathbb{R}^{B \times N \times N}$ is the batched external relation matrix. 
    Finally, convert $\tilde{\mathbf{A}}_{Hybrid}$ to its corresponding Laplacian matrix $\mathbf{L}^{lapl} \in \mathbb{R}^{B \times N \times N}$, as input to the GNN layers.
    \end{itemize}

    \item \textbf{Feature Fusion and Graph Message Passing (GCN Layers)}
    \begin{itemize}
        \item\textbf{Initial Node Feature Fusion ($\mathbf{H_0^{(0)}}$)}\\
        Before graph message passing, concatenate the features extracted from various modules to form the initial node feature vector for the GNN. These features include: multi-scale temporal features $\tilde{\mathbf{H}}^C$, local transmission risk $\mathbf{H_{ltr}}$ and global transmission risk $\mathbf{H_{gtr}}$.
        \begin{equation}
           \mathbf{H_0^{(0)}} = \text{Concat}(\tilde{\mathbf{H}}^C, \mathbf{H_{ltr}}, \mathbf{H_{gtr}}) \in \mathbb{R}^{B \times N \times (D_c + 2 hidR)}
        \end{equation} 
    
        \item\textbf{Multi-layer GCN Propagation:}\\
        Use $\mathbf{H_0^{(0)}}$ as input, propagating through several graph convolutional layers. Each graph convolutional layer aggregates neighbor information and transforms node features under the guidance of the dynamic Laplacian matrix $\mathbf{L}^{lapl}$.
        \begin{equation}
            \tilde{\mathbf{H}}^{(l+1)} = \text{LayerNorm}(\text{ReLU}(\mathbf{L}^{lapl} \tilde{\mathbf{H}}^{(l)} \tilde{\mathbf{W}}^{(l)}))
        \end{equation} 
        Here, $\tilde{\mathbf{W}}^{(l)} \in \mathbb{R}^{(D_c + 2{hidR}) \times (D_c + 2{hidR})}$ is a weight matrix, $\tilde{\mathbf{H}}^{(l)} \in \mathbb{R}^{B \times N \times (D_c + 2{hidR})}$ is the node representation at $l_{th}$ layer. $\text{LayerNorm($\cdot$)}$ is layer normalization operation, which helps stabilize training.
        The final GCN output is $\tilde{\mathbf{H}}_{GCN} \in \mathbb{R}^{B \times N \times D_{gcn}}$. If residual connections are enabled, $D_{gcn} = (n_{gcn}+1)(D_c + 2{hidR})$, where $n_{gcn}$ is the number of GCNs; otherwise, $D_{gcn} = D_c + 2{hidR}$.

    \end{itemize}
    
    \item \textbf{Prediction ($\mathbf{\hat{Y}}$)}\\
    The final prediction input consists of two concatenated parts: the final node features $\tilde{\mathbf{H}}_{GCN}$ learned by the GCN, which contain spatio-temporal information, and the pure temporal hidden states $\tilde{\mathbf{H}}_{RNN}$ extracted by the RNN. The concatenated features are fed into a linear layer, mapping them to the final prediction dimension.
    \begin{equation}
         \mathbf{res} = \text{Linear}_{output}(\text{Concat}(\tilde{\mathbf{H}}_{GCN}, \tilde{\mathbf{H}}_{RNN}))
    \end{equation} 
    $\text{Linear}_{output}$ maps $\mathbb{R}^{D_{gcn} + D_h} \to \mathbb{R}^1$, yielding $\mathbf{res} \in \mathbb{R}^{B \times N}$.
    Furthermore, if a residual window is configured, the model will also add a direct residual connection $\mathbf{P_t}$:
    \begin{equation}
         \mathbf{P_t} = \text{Linear}_{residual}(\text{Flatten}(\tilde{\mathbf{X}}_{e^{\prime}:}[:, -\text{residual\_window}:, :]))
    \end{equation} 
     The residual term $\mathbf{P}_t$ is computed by flattening the most recent $\text{residual\_window}$ time steps of input $\tilde{\mathbf{X}}_{e^{\prime}:} \in \mathbb{R}^{B \times T \times N}$ (sliced as $\mathbf{X}[:, -\text{residual\_window}:, :] \in \mathbb{R}^{B \times \text{residual\_window} \times N}$) into $\mathbb{R}^{(B \cdot N) \times \text{residual\_window}}$, then applying the linear transformation $\text{Linear}_{\text{residual}}$ to project it to $\mathbb{R}^{(B \cdot N) \times 1}$, which is ultimately reshaped into the final residual output $\mathbf{P}_t \in \mathbb{R}^{B \times N}$ for element-wise combination with the main prediction.
    
    The final prediction $\mathbf{\hat{Y}}$ is a combination of the linear layer output $\mathbf{res}$ and the residual term $\mathbf{P_t}$:
    \begin{equation}
         \mathbf{\hat{Y}} = \mathbf{res} \cdot \text{ratio} + \mathbf{P_t} \in \mathbb{R}^{B \times N}
    \end{equation} 

    where $\text{ratio}$ is an adjustable weight parameter.
    The loss function for this model remains the same as that of the ColaGNN model.
\end{itemize}

\subsection{Experiments and Analysis}
\subsubsection{Comparison with State-of-the-art}
This section is about the second set of experiments, which evaluate the performance of our proposed HybridGNN against a diverse set of state-of-the-art baseline models on four datasets: Australia-COVID, US-Regions, US-States, and Japan-Prefectures. These baselines comprise three Graph Neural Network (GNN) models (EpiGNN, ColaGNN, and STGCN), two traditional time series models (GAR and VAR), and two deep learning models (LSTM and CNNRNN-Res). We compare their prediction accuracy using MAE, RMSE, and PCC metrics, with HybridGNN demonstrating consistent superiority across these diverse benchmarks. It is important to note that for the second set of experiments, the hyperparameter settings and environmental configurations for all models (including EpiGNN, ColaGNN, and STGCN from the first experiment, as well as the newly introduced baselines) remain consistent with those established in the first experiment, ensuring a fair and comparable evaluation.

The selected baseline models are described as follows:
\begin{itemize}
    \item\textbf{EpiGNN}: As a foundational component of our proposed HybridGNN, EpiGNN is an innovative graph neural network that integrates an autoregressive (AR) component with graph convolution. 
    \item\textbf{ColaGNN}: Also a key component of our proposed HybridGNN, ColaGNN is another specialized graph neural network for epidemic prediction. It primarily leverages an attention mechanism to model complex cross-regional influences.
    \item\textbf{STGCN (Spatiotemporal Graph Convolutional Network)}: A classic and widely used spatiotemporal graph neural network. It effectively processes data with complex spatiotemporal dependencies by combining Graph Convolution with Gated Recurrent Units (GRUs).
    \item\textbf{GAR (Global Autoregression)}: A classical statistical time series model that models the current value of a variable based on its own past values, often applied globally across all regions or an aggregated series. 
    \item\textbf{VAR (Vector Autoregression)}: A classical multivariate time series statistical model. It predicts future values by establishing linear relationships between current observations and lagged observations of itself and other variables. 
    \item\textbf{LSTM (Long Short-Term Memory)}\cite{Hochreiter1997}: A widely used variant of Recurrent Neural Networks (RNNs), adept at processing and predicting long-term dependencies in time series data. 
    \item\textbf{CNNRNN-Res (Convolutional Neural Network-Recurrent Neural Network with Residual Connections)}\cite{wu2018deep}: A model that combines Convolutional Neural Networks (CNNs) and Recurrent Neural Networks (RNNs), incorporating residual connections to enhance training stability and performance. 
\end{itemize}

Next, we will proceed with a detailed analysis of the experimental results, delving into the specific performance of our proposed HybridGNN across different datasets, prediction metrics, and prediction horizons.

\begin{itemize}
    \item\textbf{Australia-COVID Dataset Prediction Analysis (daily)}

\begin{table*}[htbp]
\centering
\renewcommand{\arraystretch}{1.2}
\setlength{\tabcolsep}{3pt}
\scriptsize
\scalebox{0.8}{
\begin{tabular}{|l|l|c|c|c|c|c|c|c|c|c|c|c|}
\hline
\multicolumn{2}{|c|}{Dataset} & \multicolumn{10}{c|}{Australia horizon} \\
\hline
Method & Metric & 2 & 5 & 8 & 11 & 14 & 17 & 20 & 23 & 26 & 29 & 32 \\
\hline
\multirow{3}{*}{EpiGNN} 
  & MAE & 127.2462 & 124.4751 & 100.5464 & \underline{112.6527} & 163.1471 & 170.5775 & 194.5076 & 212.0276 & 210.8035 & 281.8333 & 304.4768\\
  & RMSE & 388.4988 & 341.1904 & 315.3492 & \underline{370.0045} & 495.6291 & 513.5239 & 573.4510 & 617.0450 & 614.6236 & 725.7447 & 828.0177 \\
  & PCC & 0.9942 & 0.9925 & 0.9919 & 0.9905 & 0.9887 & \underline{0.9864} & \underline{0.9863} & 0.9840 & 0.9810 & \underline{0.9816} & 0.9834 \\
\hline
\multirow{3}{*}{ColaGNN} 
  & MAE & 38.5352 & 68.9650 & 117.7698 & 129.8169 & 149.3325 & 154.6378 & \underline{116.4481} & 142.1623 & 145.5120 & \underline{129.1289} & \underline{108.8203}\\
  & RMSE & 180.3111 & 276.6119 & 434.2820 & 458.9208 & 523.0732 & 546.5959 & 450.7830 & 513.2499 & 547.5039 & 499.4233 & 430.2485\\
  & PCC & 0.9974 & 0.9948 & 0.9882 & 0.9878 & 0.9828 & 0.9798 & 0.9833 & 0.9798 & 0.9747 & 0.9781 & \underline{0.9841}\\
\hline
\multirow{3}{*}{STGCN} 
  & MAE & 451.3148 & 229.8330 & 295.4140 & 335.3027 & 305.1303 & 268.4486 & 679.8415 & 190.1576 & 245.7612 & 268.5856 & 413.9621\\
  & RMSE & 1052.6590 & 601.9303 & 796.3123 & 911.3586 & 842.0427 & 723.1620 & 1837.4347 & 572.1667 & 716.6028 & 683.0385 & 919.0851\\
  & PCC & 0.8412 & 0.9829 & 0.9622 & 0.9763 & 0.9835 & 0.9850 & 0.8897 & 0.9833 & 0.9818 & 0.9743 & 0.8863\\
\hline
\multirow{3}{*}{Hybrid} 
  & MAE & \underline{23.4004} & \underline{31.4465} & \textbf{59.6352} & \textbf{61.4129} & \textbf{67.5602} & \textbf{68.4927} & \textbf{74.4453} & \textbf{64.4795} & \textbf{81.6012} & \textbf{80.8418} & \textbf{75.6295}\\
  & RMSE & \underline{116.4416} & \underline{164.0043} & \textbf{222.4653} & \textbf{268.5012} & \textbf{280.7851} & \textbf{288.4950} & \textbf{322.6047} & \textbf{308.9084} & \textbf{329.5613} & \textbf{331.7577} & \textbf{337.1171}\\
  & PCC & \underline{0.9988} & \underline{0.9971} & \underline{0.9956} & \underline{0.9907} & \underline{0.9898} & \textbf{0.9893} & \textbf{0.9871} & \textbf{0.9895} & \textbf{0.9859} & \textbf{0.9859} & \textbf{0.9860}\\
\hline
\multirow{3}{*}{GAR} 
  & MAE & \textbf{8.9417} & \textbf{15.1683} & 79.1240 & 148.9641 & 264.6193 & 134.7348 & 184.2222 & 184.7187 & 184.2292 & 183.7146 & 182.5061\\
  & RMSE & \textbf{40.1244} & \textbf{69.8092} & \underline{241.3218} & 433.3076 & 750.3112 & \underline{405.9869} & 412.4550 & 417.0376 & 420.3875 & 422.8731 & 424.9285 \\
  & PCC & \textbf{0.9998} & \textbf{0.9995} & \underline{0.9985} & \textbf{0.9974} & \textbf{0.9958} & 0.9807 & 0.9800 & 0.9795 & 0.9792 & 0.9789 & 0.9787\\
\hline
\multirow{3}{*}{VAR} 
  & MAE & 157.4892 & 188.9837 & 165.8157 & 157.3335 & 150.6911 & \underline{141.9377} & 138.8578 & \underline{131.7199} & \underline{134.7356} & 137.5603 & 139.8792\\
  & RMSE & 456.6433 & 489.5019 & 489.8395 & 442.9908 & 437.2672 & 417.8038 & \underline{407.6312} & \underline{388.0964} & \underline{386.1478} & \underline{394.1534} & \underline{396.2959}\\
  & PCC & 0.9725 & 0.9679 & 0.9680 & 0.9737 & 0.9744 & 0.9767 & 0.9779 & 0.9800 & 0.9805 & 0.9802 & 0.9798\\
\hline
\multirow{3}{*}{LSTM} 
  & MAE & 32.0964 & 31.8847 & \underline{78.7997} & 211.0995 & 153.1046 & 143.2282 & 132.6313 & 522.4177 & 567.1666 & 606.9993 & 661.5128\\
  & RMSE & 169.4802 & 179.0023 & 280.9579 & 685.5695 & 563.6907 & 528.0544 & 485.7819 & 1458.4171 & 1573.6785 & 1678.5492 & 1856.1701\\
  & PCC & 0.9974 & 0.9964 & 0.9953 & 0.9776 & 0.9779 & 0.9803 & 0.9841 & \underline{0.9891} & \underline{0.9855} & 0.9796 & 0.9744\\
\hline
\multirow{3}{*}{CNNRNN-Res} 
  & MAE & 499.0875 & 121.5019 & 125.1909 & 115.6163 & \underline{122.2715} & 170.9541 & 194.8863 & 226.9638 & 240.5815 & 242.1965 & 241.9350\\
  & RMSE & 1242.8349 & 358.7626 & 377.3631 & 384.8045 & \underline{407.0471} & 515.7392 & 578.3274 & 562.5536 & 593.0843 & 599.3848 & 605.8610\\
  & PCC & 0.9559 & 0.9900 & 0.9883 & 0.9872 & 0.9856 & 0.9838 & 0.9838 & 0.9753 & 0.9743 & 0.9741 & 0.9751\\
\hline
\end{tabular}
}
\caption{Performance Comparison of All Prediction Models on Australia-COVID Dataset with Horizon = 2, 5, 8, 11, 14, 17, 20, 23, 26, 29, 32. The underlined indicates the second-best, and bold indicates the best.}
\label{tab:comparison between all models on Australia dataset}
\end{table*}

\begin{figure}[H]
        \centering
        \begin{minipage}[b]{0.48\textwidth} 
            \includegraphics[width=\linewidth]{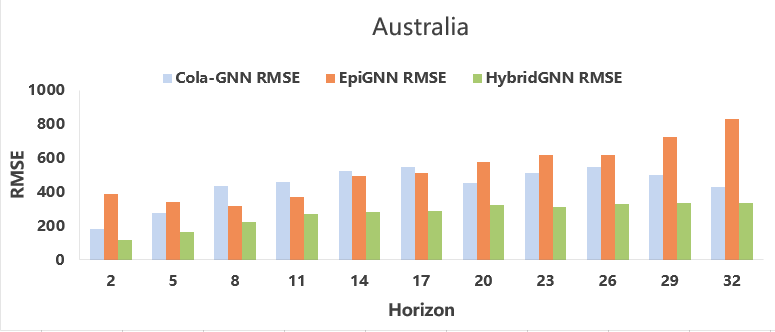}
        \end{minipage}
        \hfill
        \begin{minipage}[b]{0.48\textwidth} 
            \includegraphics[width=\linewidth]{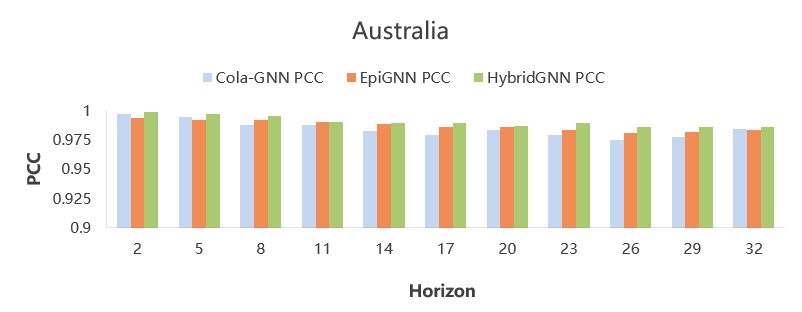}
        \end{minipage}
        \caption{Performance Comparison between HybridGNN and two baselines (EpiGNN and ColaGNN) on Australia-COVID dataset}
        \label{fig:comparison between Hybrid and two baselines on Australia-COVID}
    \end{figure}

The HybridGNN model demonstrates outstanding performance on the Australia dataset. Compared to the fusion models EpiGNN and ColaGNN, HybridGNN consistently achieves lower MAE and RMSE values, along with higher PCC scores, indicating superior prediction accuracy and stronger correlation with real-world data, which is illustrated in Figure \ref{fig:comparison between Hybrid and two baselines on Australia-COVID}. Furthermore, when measured against all other models in the Table \ref{tab:comparison between all models on Australia dataset} (including STGCN, GAR, VAR, LSTM, and CNNRNN-Res), HybridGNN maintains the lowest MAE and RMSE values across almost all horizons, while also securing the leading position in PCC. This unequivocally highlights HybridGNN's exceptional predictive capabilities and stability on the Australia dataset.

\begin{table}[htbp]
\centering
\label{tab:results}
\resizebox{\textwidth}{!}{%
\begin{tabular}{l l c c c c c c c c c c c c c c c}
\toprule
\multirow{2}{*}{Dataset} & \multirow{2}{*}{} & \multicolumn{5}{c}{Japan} & \multicolumn{5}{c}{US-Regions} & \multicolumn{5}{c}{US-States} \\
\cmidrule(lr){3-7} \cmidrule(lr){8-12} \cmidrule(lr){13-17}
 & & \multicolumn{5}{c}{horizon} & \multicolumn{5}{c}{horizon} & \multicolumn{5}{c}{horizon} \\
\cmidrule(lr){3-7} \cmidrule(lr){8-12} \cmidrule(lr){13-17}
Method & Metric & 2 & 5 & 14 & 23 & 29 & 2 & 5 & 14 & 23 & 29 & 2 & 5 & 14 & 23 & 29 \\
\midrule
EpiGNN & MAE & 317.7547 & 421.4113 & 532.3716 & 580.0948 & 572.4716 & 271.9734 & 543.7277 & \textbf{584.1873} & 743.6681 & 679.3557 & 58.6324 & 85.646 & 107.6102 & 108.9413 & 138.4842 \\
 & RMSE & \underline{934.0154} & \textbf{1051.202} & 1466.451 & 1403.616 & 1551.241 & \underline{495.2303} & \underline{898.0375} & \textbf{994.9219} & 1203.012 & 1043.007 & 148.2381 & \underline{199.371} & \underline{228.3557} & \underline{236.3079} & 265.86 \\
 & PCC & 0.9168 & \textbf{0.9016} & 0.7321 & 0.7966 & 0.722 & 0.9421 & 0.7813 & \textbf{0.7312} & 0.5585 & 0.6884 & 0.9449 & \underline{0.8975} & 0.8601 & 0.8508 & 0.8088 \\
\midrule
ColaGNN & MAE & 287.5924 & \underline{379.4594} & \underline{491.446} & \underline{499.7023} & \underline{492.9672} & \underline{233.6242} & 525.8035 & 708.7531 & \underline{658.7186} & \underline{484.4075} & 53.4726 & 83.3606 & \textbf{95.2117} & \underline{101.7764} & \underline{102.4193} \\
 & RMSE & 948.7436 & 1152.288 & \underline{1412.874} & \underline{1383.584} & 1540.414 & 506.9187 & 1007.584 & 1260.498 & \underline{1061.498} & \underline{969.8692} & 148.1823 & 221.7605 & 229.4957 & 238.5425 & \underline{233.7741} \\
 & PCC & \textbf{0.9253} & 0.8628 & \underline{0.7812} & \underline{0.8098} & 0.6967 & 0.9435 & 0.7988 & \underline{0.6874} & \underline{0.6701} & \underline{0.774} & \underline{0.9498} & 0.8747 & \textbf{0.8797} & \underline{0.8663} & \underline{0.8613} \\
\midrule
STGCN & MAE & 304.3613 & 433.1643 & 538.246 & 568.3075 & 518.1366 & 408.5528 & 556.1506 & 800.3231 & 768.8422 & 587.5015 & 82.7078 & 101.3556 & 112.4669 & 111.9671 & 108.3811 \\
 & RMSE & 967.6338 & 1301.511 & 1598.077 & 1434.441 & \underline{1455.733} & 697.2763 & 1066.68 & 1381.32 & 1305.12 & 1077.738 & 181.8974 & 235.4713 & 271.1715 & 254.2331 & 244.7531 \\
 & PCC & 0.9071 & 0.8382 & 0.7752 & 0.7725 & \textbf{0.8322} & 0.8783 & 0.7682 & 0.4352 & 0.5499 & 0.7464 & 0.914 & 0.8517 & 0.8092 & 0.8281 & 0.8453 \\
\midrule
HybridGNN & MAE & \textbf{270.804} & \textbf{379.3438} & \textbf{486.6925} & \textbf{470.0945}& \textbf{479.2033} & \textbf{227.9353} & \textbf{434.3308} & \underline{658.3672} & \textbf{590.7435} & \textbf{449.7577} & \textbf{52.0503} & \textbf{76.3155} & \underline{100.0204} & \textbf{99.3903} & \textbf{101.9999} \\
 & RMSE & \textbf{921.458} & \underline{1121.055} & \textbf{1408.038} & \textbf{1319.952} & \textbf{1410.298} & \textbf{493.3472} & \textbf{833.6702} & \underline{1157.178} & \textbf{993.2136} & \textbf{890.3559} & \textbf{137.7529} & \textbf{196.3588} & \textbf{225.7914} & \textbf{220.7136} & \textbf{226.6243} \\
 & PCC & \underline{0.9217} & \underline{0.8847} & \textbf{0.7848} & \textbf{0.8252} & \underline{0.7812} & \textbf{0.9464} & \textbf{0.8321} & 0.6504 & \textbf{0.7408} & \textbf{0.7988} & \textbf{0.9547} & \textbf{0.9058} & \underline{0.8663} & \textbf{0.8764} & \textbf{0.8713} \\
\midrule
GAR & MAE & 348.4744 & 636.3506 & 698.4774 & 698.1638 & 598.0141 & 261.3882 & 525.5662 & 848.754 & 750.9815 & 552.1652 & 55.9509 & 93.999 & 144.8457 & 138.0241 & 120.9852 \\
 & RMSE & 1229.896 & 1976 & 2021.932 & 2029.766 & 1726.18 & 542.1134 & 1000.35 & 1488.593 & 1338.422 & 1027.191 & 149.855 & 234.4273 & 340.4244 & 330.7043 & 285.3659 \\
 & PCC & 0.8049 & 0.3494 & 0.4629 & 0.4532 & 0.5547 & 0.9319 & 0.7919 & 0.5115 & 0.5285 & 0.7468 & 0.9438 & 0.8745 & 0.7435 & 0.7458 & 0.7911 \\
\midrule
VAR & MAE & 511.1963 & 770.097 & 746.0118 & 802.3989 & 676.3038 & 420.2119 & 626.335 & 828.1846 & 784.1926 & 638.3933 & 111.8088 & 130.3733 & 160.3149 & 179.0292 & 153.7718 \\
 & RMSE & 1365.755 & 2032.304 & 1900.42 & 1895.832 & 1714.732 & 727.6051 & 1047.776 & 1318.061 & 1264.515 & 1081.703 & 228.0389 & 279.3023 & 337.2174 & 382.3912 & 320.2148 \\
 & PCC & 0.7496 & 0.2849 & 0.4814 & 0.45 & 0.564 & 0.683 & 0.6942 & 0.4577 & 0.5543 & 0.6698 & 0.8619 & 0.7884 & 0.6994 & 0.6216 & 0.7252 \\
\midrule
LSTM & MAE & \underline{274.2626} & 399.1941 & 536.7787 & 551.9538 & 510.3136 & 243.4456 & \underline{485.064} & 760.6742 & 736.6856 & 741.2546 & \underline{52.3984} & \underline{78.9112} & 125.9012 & 145.2799 & 147.0832 \\
 & RMSE & 968.1576 & 1181.527 & 1599.631 & 1390.769 & 1530.291 & 528.9254 & 972.9679 & 1231.165 & 1204.848 & 1206.959 & \underline{144.6276} & 207.0022 & 300.834 & 341.6578 & 343.9574 \\
 & PCC & 0.9037 & 0.8614 & 0.7312 & 0.7525 & 0.7573 & 0.9397 & \underline{0.8207} & 0.5282 & 0.5659 & 0.5626 & 0.948 & 0.8932 & 0.7619 & 0.7328 & 0.7301 \\
\midrule
CNNRNN-Res & MAE & 337.4548 & 661.3301 & 745.3217 & 910.321 & 617.5538 & 317.7955 & 499.8897 & 795.5295 & 720.0831 & 855.8668 & 93.2001 & 107.1022 & 111.1416 & 108.3219 & 108.1891 \\
 & RMSE & 1063.797 & 1984.306 & 2110.505 & 1960.829 & 1672.761 & 588.6313 & 926.6054 & 1299.886 & 1257.29 & 1483.717 & 215.2229 & 258.2973 & 262.0959 & 273.2706 & 258.789 \\
 & PCC & 0.8606 & 0.3336 & 0.1309 & 0.3112 & 0.6858 & 0.9177 & 0.787 & 0.4575 & 0.5504 & 0.4687 & 0.8817 & 0.8343 & 0.821 & 0.8429 & 0.8436 \\
\bottomrule
\end{tabular}
}
\caption{Performance comparison across three datasets with representative horizons: short-term (2,5), mid-term (14), and long-term (23,29)}
\label{tab:comparison between all models across three datasets}
\end{table}

\item\textbf{US-Regions Dataset Prediction Analysis (weekly)}

\begin{figure}[H]
        \centering
        \begin{minipage}[b]{0.48\textwidth} 
            \includegraphics[width=\linewidth]{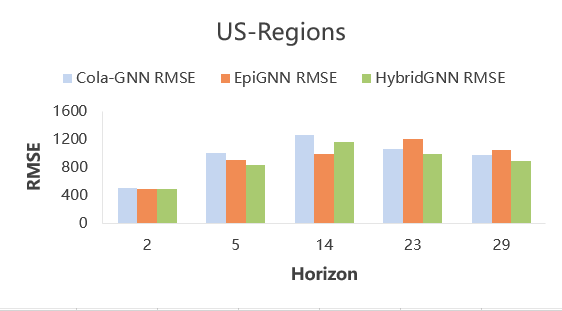}
        \end{minipage}
        \hfill
        \begin{minipage}[b]{0.48\textwidth} 
            \includegraphics[width=\linewidth]{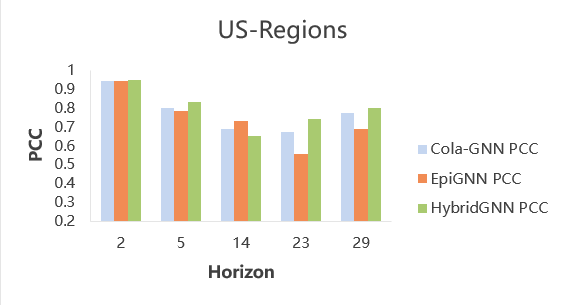}
        \end{minipage}
        \caption{Performance Comparison between HybridGNN and two baselines (EpiGNN and ColaGNN) on US-Regions dataset}
        \label{fig:comparison between Hybrid and two baselines on US-Regions}
    \end{figure}
    
On the US-Regions dataset, the HybridGNN model demonstrates significant performance advantages, as is shown in Table \ref{tab:comparison between all models across three datasets} and further illustrated in Figure \ref{fig:comparison between Hybrid and two baselines on US-Regions}. Compared to the fusion models EpiGNN and ColaGNN, HybridGNN consistently achieves lower MAE and RMSE values across most prediction horizons, and generally higher PCC values. This indicates its predictions are more accurate and show a stronger correlation with actual data, although EpiGNN was slightly superior in MAE and RMSE at horizon 14, and ColaGNN also slightly better in MAE at that same horizon. When considering all models, HybridGNN largely stands out with the lowest MAE and RMSE values and leading PCC scores across various time steps, fully proving its excellent predictive performance and stability on the US-Regions dataset.

\item\textbf{US-States Dataset Prediction Analysis (weekly)}

\begin{figure}[H]
        \centering
        \begin{minipage}[b]{0.48\textwidth} 
            \includegraphics[width=\linewidth]{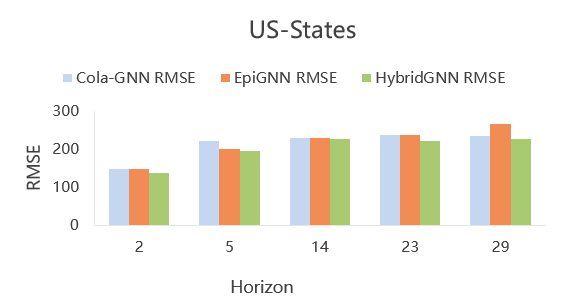}
        \end{minipage}
        \hfill
        \begin{minipage}[b]{0.48\textwidth} 
            \includegraphics[width=\linewidth]{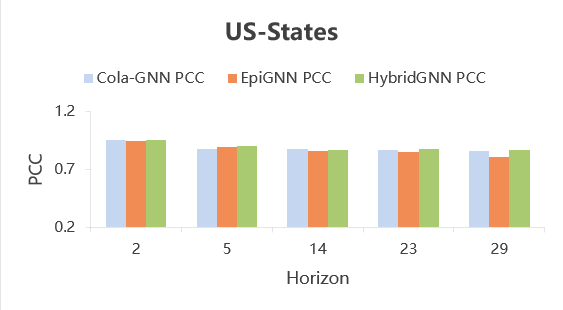}
        \end{minipage}
        \caption{Performance Comparison between HybridGNN and two baselines (EpiGNN and ColaGNN) on US-States dataset}
        \label{fig:comparison between Hybrid and two baselines on US-States}
    \end{figure}

On the US-States dataset, the HybridGNN model demonstrates consistently superior overall performance, which is detailed in Table \ref{tab:comparison between all models across three datasets} and further illustrated in Figure \ref{fig:comparison between Hybrid and two baselines on US-States}. It achieves the lowest RMSE values across all time horizons, significantly outperforming all comparative models, including the fusion models EpiGNN and ColaGNN. This indicates its predictions have the smallest error margins. For MAE and PCC, HybridGNN also generally performs best across most prediction periods, yielding the lowest MAE and highest PCC values. The only exception is at horizon 14, where ColaGNN shows slightly better performance in both MAE and PCC. Overall, HybridGNN's notable advantages in error control and data correlation prove its strong predictive capabilities on the US-States dataset.

\item\textbf{Japan-Prefectures Dataset Prediction Analysis (weekly)}

\begin{figure}[H]
        \centering
        \begin{minipage}[b]{0.48\textwidth} 
            \includegraphics[width=\linewidth]{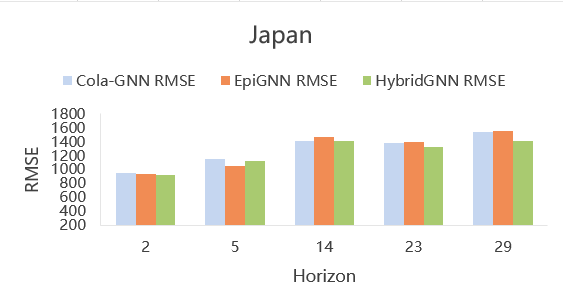}
        \end{minipage}
        \hfill
        \begin{minipage}[b]{0.48\textwidth} 
            \includegraphics[width=\linewidth]{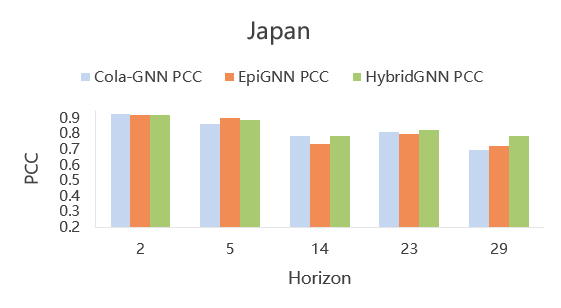}
        \end{minipage}
        \caption{Performance Comparison between HybridGNN and two baselines (EpiGNN and ColaGNN) on Japan-Prefectures dataset}
        \label{fig:comparison between Hybrid and two baselines on Japan-Prefectures}
    \end{figure}

The analysis is detailed in Table \ref{tab:comparison between all models across three datasets} and further illustrated in Figure \ref{fig:comparison between Hybrid and two baselines on Japan-Prefectures}. On the Japan-Prefectures dataset, the HybridGNN model demonstrates remarkably superior overall performance. Compared to the fusion models EpiGNN and ColaGNN, HybridGNN achieves the best MAE across all time horizons and leads in RMSE for most horizons, indicating significantly higher prediction accuracy than both. While ColaGNN and EpiGNN show slightly better PCC in some short-term predictions (e.g., horizons 2 and 5), HybridGNN performs best in PCC for mid-to-long-term forecasts (horizons 14 and 23). When compared against all other models, HybridGNN's universal leadership in MAE and its majority advantage in RMSE highlight its robust error control capabilities, making it the most outstanding model on this dataset, with other models (like STGCN at horizon 29) only marginally excelling in isolated PCC metrics.
\end{itemize}

\textbf{Summary:}
Overall, our proposed HybridGNN demonstrates significant and often superior performance across the evaluated datasets. Its superiority is most pronounced on the Australia-COVID and US-States datasets, where it consistently achieves the best RMSE and PCC across almost all prediction horizons. This demonstrates the powerful synergy achieved by combining the complementary strengths of EpiGNN and ColaGNN.
On the US-Regions dataset, HybridGNN again proves to be the most robust model, delivering the lowest error in the crucial long-term stages and effectively balancing short-term accuracy with long-term stability.
Crucially, on the highly volatile Japan-Prefectures dataset, HybridGNN delivers a highly competitive and often superior performance. It secures the lowest error at multiple key horizons, showcasing its robustness and adaptability even in complex, high-volatility scenarios.
In conclusion, HybridGNN establishes itself not just as a general and robust model, but as a top-tier performer for epidemic forecasting, demonstrating its potential by excelling in both stable and highly challenging environments.

\subsubsection{Ablation Tests}

To analyze the effect of each component in our framework, we perform ablation tests with the following settings:
        
\begin{itemize}
    \item \textbf{Hybrid w/o ctemp}: Removes \texttt{Conv1D} to capture short-term and long-term temporal dependencies, but instead uses a simple linear layer to process the raw time-series input as node features.
    \item \textbf{Hybrid w/o LTR}: Removes the local transmission risk encoding module. This means that the node feature $\mathbf{H}_{ltr}$ will not be concatenated into the GNN's input during the feature fusion stage.
    \item \textbf{Hybrid w/o GTR}: Removes the global transmission risk encoding module. This means that the node feature $\mathbf{H}_{gtr}$ will not be concatenated into the GNN's input during the feature fusion stage.
    \item \textbf{Hybrid w/o dygraph}: Removes ColaGNN-style dynamic graph. In this case, the model only uses the static geographical adjacency matrix combined with degree information to construct the final adjacency matrix, without employing the RNN-based dynamic attention mechanism.

\end{itemize}
\begin{table}[H]
    \centering
    \scalebox{0.55}{
    \begin{tabular}{lccccccccccc}
        \toprule
        RMSE & \textbf{2} & \textbf{5} & \textbf{8} & \textbf{11} & \textbf{14} & \textbf{17} & \textbf{20} & \textbf{23} & \textbf{26} & \textbf{29} & \textbf{32} \\
        \midrule
        \multicolumn{12}{c}{\textit{Australia}} \\
        \midrule
        Hybrid w/o ctemp & 415.9470 & 1991.1867 & 11594.0336 & 433.4485 & 1916.5805 & 1931.0230 & 1956.7124 & 1974.4552 & 1874.6233 & 1902.2566 & 1975.7156 \\
        Hybrid w/o LTR & 273.6972 & 356.7588 & 381.4664 & 451.2182 & 696.5153 & 1178.9851 & 1138.2398 & 1418.4840 & 1385.6420 & 1506.8618 & 1595.1749 \\
        Hybrid w/o GTR & 243.9937 & 287.4677 & 356.9107 & 326.7577 & 820.5287 & 964.1199 & 1145.3360 & 1393.4021 & 1450.5515 & 1528.8793 & 1591.1614 \\
        Hybrid w/o dygraph & 1912.3533 & 1961.3242 & 1998.8525 & 2032.4530 & 2036.5009 & 2034.0261 & 2032.8792 & 2006.7390 & 2010.6395 & 2019.1537 & 2028.1556 \\
        Hybrid & \textbf{116.4416} & \textbf{164.0043} & \textbf{222.4653} & \textbf{268.5012} & \textbf{280.7851} & \textbf{288.4950} & \textbf{322.6047} & \textbf{308.9084} & \textbf{329.5613} & \textbf{331.7577} & \textbf{337.1171} \\
        \midrule
        \multicolumn{12}{c}{\textit{US-States}} \\
        \midrule
        Hybrid w/o ctemp & 147.8191 & 218.4046 & 232.9632 & 270.1645 & 243.2812 & 322.0208 & 332.0216 & 292.4265 & 271.2481 & 252.0407 & 248.6336 \\
        Hybrid w/o LTR & 144.2372 & 216.5336 & 239.4663 & 252.3514 & 227.7764 & 245.1995 & 260.3577 & 289.3448 & \textbf{224.8867} & 254.3987 & 237.0310 \\
        Hybrid w/o GTR & 141.4485 & 206.0671 & 232.3240 & 241.0928 & 229.4291 & 242.5869 & 246.6565 & 248.1291 & 261.1806 & 246.2272 & 246.9301 \\
        Hybrid w/o dygraph & 191.5346 & 227.8267 & 253.8217 & 254.7951 & 252.8139 & 251.5027 & 246.1103 & 242.6997 & 248.6554 & 256.4419 & 254.6796 \\
        Hybrid & \textbf{137.7529} & \textbf{196.3538} & \textbf{231.9247} & \textbf{235.4558} & \textbf{225.7914} & \textbf{235.5321} & \textbf{220.5349} & \textbf{220.7136} & 231.0958 & \textbf{226.6243} & \textbf{234.6385} \\
        \bottomrule
    \end{tabular}
    }
    \caption{RMSE Performance of HybridGNN Variants in Ablation Study on Australia and US-States Datasets (Lower is better)}
    \label{tab:rmse}
\end{table}
The results for RMSE are shown in the Table~\ref{tab:rmse}. We observe that:\\
\begin{itemize}
    \item \textbf{Overall Trend}: In most cases, the full model (Hybrid) performs optimally in terms of RMSE, especially on the Australia dataset, where its RMSE values are significantly lower than all ablation variants. This indicates that all components work collaboratively to effectively reduce prediction error.
    \item \textbf{Importance of Temporal Features (Hybrid vs Hybrid w/o ctemp)}: On the Australia dataset, removing temporal features (Hybrid w/o ctemp) leads to a significant increase in RMSE, particularly for short-term predictions (e.g., 2, 5, 8, 11), sometimes by orders of magnitude worse than other ablated models. This highlights the critical role of temporal feature extraction in capturing Australian influenza trends. On the US-States dataset, removing temporal features also generally increases RMSE, though not as drastically as for Australia.
    \item \textbf{Importance of Global Transmission Risk (Hybrid vs Hybrid w/o GTR)}: Removing GTR (Hybrid w/o GTR) generally leads to an increase in RMSE. On the Australia dataset, the presence of GTR significantly reduces RMSE for both short-term and long-term predictions. On the US-States dataset, GTR also plays a key role in reducing RMSE at multiple time steps (e.g., 11, 17).
    \item \textbf{Importance of Local Transmission Risk (Hybrid vs Hybrid w/o LTR)}: Removing LTR (Hybrid w/o LTR) also generally leads to an increase in RMSE. On the Australia dataset, LTR's contribution to RMSE reduction is similar to GTR's. On the US-States dataset, LTR has a noticeable impact on some short-term and long-term predictions (e.g., 2, 23, 32), but is not as universally beneficial as GTR.
    \item \textbf{Importance of Dynamic Graph (Hybrid vs Hybrid w/o dygraph)}: Removing the dynamic graph (Hybrid w/o dygraph) leads to a substantial degradation in RMSE on the Australia dataset, indicating that the dynamic graph is crucial for capturing complex inter-regional relationships. On the US-States dataset, while also degraded, the magnitude is less severe than for Australia, which might suggest the initial importance of geographical structure in the US dataset.
\end{itemize}

\begin{table}[H]
    \centering
    \scalebox{0.65}{
    \begin{tabular}{lccccccccccc}
        \toprule
        PCC & \textbf{2} & \textbf{5} & \textbf{8} & \textbf{11} & \textbf{14} & \textbf{17} & \textbf{20} & \textbf{23} & \textbf{26} & \textbf{29} & \textbf{32} \\
        \midrule
        \multicolumn{12}{c}{\textit{Australia}} \\
        \midrule
        Hybrid w/o ctemp & 0.9859 & 0.2674 & 0.0291 & 0.9898 & 0.3228 & 0.2412 & 0.2461 & 0.1803 & 0.3199 & 0.3595 & 0.1857 \\
        Hybrid w/o LTR & 0.9945 & 0.9935 & 0.9924 & 0.9910 & 0.9890 & 0.9837 & 0.9847 & 0.9795 & 0.9816 & 0.9757 & 0.9753 \\
        Hybrid w/o GTR & 0.9959 & 0.9956 & 0.9936 & 0.9883 & \textbf{0.9900} & 0.9876 & 0.9852 & 0.9840 & 0.9807 & 0.9796 & 0.9772 \\
        Hybrid w/o dygraph & 0.3523 & 0.3686 & 0.3477 & 0.2833 & 0.2892 & 0.3593 & 0.3496 & 0.3780 & 0.3916 & 0.3848 & 0.3577 \\
        Hybrid & \textbf{0.9988} & \textbf{0.9971} & \textbf{0.9956} & \textbf{0.9950} & 0.9898 & \textbf{0.9893} & \textbf{0.9871} & \textbf{0.9895} & \textbf{0.9859} & \textbf{0.9856} & \textbf{0.9860} \\
        \midrule
        \multicolumn{12}{c}{\textit{US-States}} \\
        \midrule
        Hybrid w/o ctemp & 0.9485 & 0.8847 & 0.8741 & 0.8185 & 0.8643 & 0.7640 & 0.7220 & 0.7900 & 0.8129 & 0.8302 & 0.8385 \\
        Hybrid w/o LTR & 0.9526 & 0.8891 & 0.8536 & 0.8417 & 0.8672 & 0.8557 & 0.8594 & 0.8267 & 0.8716 & 0.8404 & 0.8495 \\
        Hybrid w/o GTR & 0.9528 & 0.9033 & \textbf{0.8752} & 0.8626 & \textbf{0.8779} & \textbf{0.8690} & 0.8596 & 0.8663 & 0.8474 & 0.8434 & 0.8416 \\
        Hybrid w/o dygraph & 0.9039 & 0.8630 & 0.8319 & 0.8356 & 0.8321 & 0.8467 & 0.8509 & 0.8452 & 0.8439 & 0.8314 & 0.8271 \\
        Hybrid & \textbf{0.9547} & \textbf{0.9058} & 0.8633 & \textbf{0.8635} & 0.8663 & 0.8511 & \textbf{0.8730} & \textbf{0.8746} & \textbf{0.8746} & \textbf{0.8713} & \textbf{0.8505} \\
        \bottomrule
        \end{tabular}
    }
    \caption{PCC Performance of HybridGNN Variants in Ablation Study on Australia and US-States Datasets (Higher is better)}
    \label{tab:pcc}
\end{table}
The results for PCC are shown in the Table~\ref{tab:pcc}. We observe that:
\begin{itemize}
    \item \textbf{Overall Trend}: The full model (Hybrid) performs best in terms of PCC, especially on the Australia dataset, where its PCC values are generally the highest, close to 1. This indicates that the complete model excels at capturing the linear relationship between predictions and true values.
    \item \textbf{Importance of Temporal Features (Hybrid vs Hybrid w/o ctemp)}: Removing temporal features (Hybrid w/o ctemp) leads to a significant drop in PCC. On the Australia dataset, some PCC values become very low (close to 0), demonstrating a loss of ability to capture sequence trends. This again emphasizes the necessity of temporal feature extraction.
    \item \textbf{Importance of Global Transmission Risk (Hybrid vs Hybrid w/o GTR)}: Removing GTR (Hybrid w/o GTR) generally leads to a decrease in PCC. On the Australia dataset, although it performs sub-optimally at certain time steps (e.g., 8, 11, 14), overall it does not match the performance of the full model. On the US-States dataset, GTR's contribution to PCC improvement is significant at several time steps (e.g., 8, 11, 14, 17).
    \item \textbf{Importance of Local Transmission Risk (Hybrid vs Hybrid w/o LTR)}: Removing LTR (Hybrid w/o LTR) also generally leads to a decrease in PCC. On the Australia dataset, LTR's contribution to PCC is similar to GTR's. On the US-States dataset, LTR performs exceptionally well at some time steps (e.g., 32), even outperforming the full model, but is less effective at others.
    \item \textbf{Importance of Dynamic Graph (Hybrid vs Hybrid w/o dygraph)}: Removing the dynamic graph (Hybrid w/o dygraph) results in a significant drop in PCC on both datasets, especially on the Australia dataset where PCC values become very low, reiterating the crucial role of the dynamic graph in modeling complex spatial dependencies.
\end{itemize}

In summary, the table data strongly demonstrates the importance of temporal feature extraction, global transmission risk encoding, local transmission risk encoding, and dynamic graph construction components within the \texttt{HybridGNN} model for improving prediction accuracy and correlation. Specifically, temporal features and the dynamic graph are crucial for the model's robustness and performance, and these modules show greater advantage in long-term predictions.

\subsubsection{Sensitivity Analysis}
\begin{table}[H]
    \centering
    \scalebox{0.8}{
    \begin{tabular}{ccccc}
        \toprule
        \textbf{Look-back Window} & \textbf{US-States} & \textbf{Australia} & \textbf{US-Regions} & \textbf{Japan} \\
        \cmidrule(lr){1-1} \cmidrule(lr){2-2} \cmidrule(lr){3-3} \cmidrule(lr){4-4} \cmidrule(lr){5-5}
        & \textbf{RMSE} / \textbf{PCC} & \textbf{RMSE} / \textbf{PCC} & \textbf{RMSE} / \textbf{PCC} & \textbf{RMSE} / \textbf{PCC} \\
        \midrule
        10 & 328.5342 / 0.7433 & 901.5119 / 0.9902 & 1266.6971 / 0.4745 & 1791.2859 / 0.6202 \\
        20 & 237.8147 / 0.8458 & \textbf{298.4658} / 0.9894 & 1431.4211 / 0.4757 & 1539.2149 / 0.7511 \\
        30 & 235.0392 / \textbf{0.8836} & 546.33 / 0.9913 & 1000.9503 / \textbf{0.783} & 1749.5062 / 0.6332 \\
        40 & 251.5761 / 0.8702 & 452.632 / \textbf{0.9915} & \textbf{963.2282} / 0.74 & \textbf{1399.8574} / \textbf{0.8072} \\
        50 & \textbf{225.6077} / 0.8765 & 842.7432 / 0.9899 & 1107.5108 / 0.6713 & 1766.8719 / 0.7078 \\
        \bottomrule
    \end{tabular}
    }
    \caption{Performance Metrics (RMSE and PCC) for Different Look-back Windows (horizon=15)}
    \label{tab:lookback_window}
\end{table}

\begin{table}[H]
    \centering
    \scalebox{0.85}{
    \begin{tabular}{ccccc}
        \toprule
        \textbf{\#Filter} & \textbf{US-States} & \textbf{Australia} & \textbf{US-Regions} & \textbf{Japan} \\
        \cmidrule(lr){1-1} \cmidrule(lr){2-2} \cmidrule(lr){3-3} \cmidrule(lr){4-4} \cmidrule(lr){5-5}
        & \textbf{MAE} / \textbf{RMSE} & \textbf{MAE} / \textbf{RMSE} & \textbf{MAE} / \textbf{RMSE} & \textbf{MAE} / \textbf{RMSE} \\
        \midrule
        4 & 115.8689 / 257.4315 & 175.8813 / 514.0602 & \textbf{678.907} / \textbf{1235.6178} & 594.0095 / 1701.2319 \\
        8 & 107.9496 / 237.9043 & \textbf{73.2646} / \textbf{280.7642} & 813.9878 / 1431.4215 & 585.1898 / \textbf{1580.785} \\
        16 & \textbf{101.8772} / \textbf{237.2356} & 173.789 / 529.2308 & 765.532 / 1353.777 & 770.1307 / 1689.6497 \\
        32 & 134.9709 / 313.0276 & 295.3007 / 836.9718 & 783.0477 / 1326.0179 & \textbf{576.2716} / 1710.5112 \\
        64 & 108.1324 / 254.8871 & 196.1771 / 562.5098 & 782.9492 / 1337.6009 & 604.6296 / 1705.8411 \\
        \bottomrule
    \end{tabular}
    }
    \caption{Performance Metrics (MAE and RMSE) for Different Filter Numbers (horizon=15)}
    \label{tab:filter_number}
\end{table}

\begin{table}[H]
    \centering
    \scalebox{0.8}{
    \begin{tabular}{ccccc}
        \toprule
        \textbf{Learning Rate} & \textbf{US-States} & \textbf{Australia} & \textbf{US-Regions} & \textbf{Japan} \\
        \cmidrule(lr){1-1} \cmidrule(lr){2-2} \cmidrule(lr){3-3} \cmidrule(lr){4-4} \cmidrule(lr){5-5}
        & \textbf{MAE} / \textbf{RMSE} & \textbf{MAE} / \textbf{RMSE} & \textbf{MAE} / \textbf{RMSE} & \textbf{MAE} / \textbf{RMSE} \\
        \midrule
        0.001 & 107.9496 / 237.9043 & 73.2646 / 280.7642 & 813.9878 / 1431.4215 & 585.1898 / 1580.785 \\
        0.005 & \textbf{98.1949} / \textbf{223.8596} & 70.1162 / \textbf{275.4259} & \textbf{748.0739} / \textbf{1250.8788} & 518.6663 / 1513.5086 \\
        0.01 & 113.3158 / 276.8495 & \textbf{55.2511} / 276.8183 & 761.1348 / 1305.3237 & \textbf{511.2254} / \textbf{1457.708} \\
        \bottomrule
    \end{tabular}
    }
    \caption{Performance Metrics (MAE and RMSE) for Different Learning Rates (horizon=15)}
    \label{tab:learning_rate}
\end{table}

\begin{table}[H]
    \centering
    \scalebox{0.8}{
    \begin{tabular}{ccccc}
        \toprule
        \textbf{RNN Dimension} & \textbf{US-States} & \textbf{Australia} & \textbf{US-Regions} & \textbf{Japan} \\
        \cmidrule(lr){1-1} \cmidrule(lr){2-2} \cmidrule(lr){3-3} \cmidrule(lr){4-4} \cmidrule(lr){5-5}
        & \textbf{MAE} / \textbf{RMSE} & \textbf{MAE} / \textbf{RMSE} & \textbf{MAE} / \textbf{RMSE} & \textbf{MAE} / \textbf{RMSE} \\
        \midrule
        10 & 135.515 / 300.9511 & 293.4158 / 830.2479 & 788.5283 / 1381.8195 & 666.9337 / 1860.412 \\
        20 & 109.5947 / 239.8612 & 68.1439 / 300.171 & 813.9877 / 1431.4214 & 599.566 / 1606.8013 \\
        30 & \textbf{105.7395} / 235.3523 & 88.8185 / 365.8249 & \textbf{722.9141} / \textbf{1265.7188} & 778.213 / 1749.8347 \\
        40 & 118.7727 / 263.5005 & 90.3212 / \textbf{291.4734} & 762.4515 / 1325.7405 & 574.501 / 1620.651 \\
        50 & 106.0453 / \textbf{233.0153} & \textbf{67.6961} / 313.1836 & 738.3013 / 1319.2782 & \textbf{565.5046} / \textbf{1601.9304} \\
        \bottomrule
    \end{tabular}
    }
    \caption{Performance Metrics (MAE and RMSE) for Different RNN Dimensions (horizon=15)}
    \label{tab:rnn_dimension}
\end{table}

This section evaluates the robustness of the model’s performance with respect to key hyperparameters, including the look-back window, number of filters, learning rate, and RNN dimension. The analysis leverages RMSE, PCC (for look-back window experiments), and MAE (for other cases) as primary metrics, applied to four distinct datasets: US-States, Australia-COVID, US-Regions, and Japan-Prefectures. All experiments maintain a consistent horizon of 15.

\begin{itemize}
\item \textbf{Variation in Look-back Window}
To evaluate the model’s reliance on the extent of historical data, we tested look-back windows at intervals of 10, 20, 30, 40, and 50. In Table \ref{tab:lookback_window}, for US-States, RMSE fluctuates between 225.6077 and 328.5342, with the lowest value at 50, and PCC ranges from 0.7433 to 0.8836, peaking at 30, suggesting a degree of resilience to window size changes. Australia displays a wider RMSE spread (298.4658 to 901.5119), with the best outcome at 20, and PCC values remaining consistently high (0.9894 to 0.9915), pointing to notable sensitivity. US-Regions achieves its minimum RMSE at 40 (963.2282, PCC 0.74), while Japan optimizes at 40 (RMSE 1399.8574, PCC 0.8072), with a moderate range. The significant RMSE variation in Australia (603.0461) suggests a strong influence from this parameter, likely tied to underlying data characteristics.

\item \textbf{Adjustment of Filter Count}
The model incorporates a convolutional layer to process time series data, where the number of filters plays a role in feature extraction. We experimented with filter counts of 4, 8, 16, 32, and 64, with performance metrics detailed in Table \ref{tab:filter_number}. US-States records its lowest RMSE at 16 filters (237.2356, MAE 101.8772), while Australia performs best at 8 (RMSE 280.7642, MAE 73.2646). US-Regions shows an optimal RMSE at 4 filters (1235.6178, MAE 678.907), and Japan at 32 (RMSE 1580.785, MAE 576.2716). The RMSE range is most pronounced for Australia (556.2076), indicating high sensitivity, whereas Japan exhibits the smallest variation (129.7262). Performance tends to decline with excessive filters (e.g., 64), hinting at potential overfitting with larger configurations.

\item \textbf{Modification of Learning Rate}
The learning rate, a fundamental aspect of the training process, was varied across 0.001, 0.005, and 0.01 to assess its impact. Results, presented in Table \ref{tab:learning_rate}, show US-States achieving the best RMSE at 0.005 (223.8596, MAE 98.1949), with a sensitivity range of 52.9899. Australia displays minimal fluctuation (RMSE 275.4259 to 280.7642), optimizing at 0.01 (RMSE 276.8183, MAE 55.2511). US-Regions peaks at 0.005 (RMSE 1250.8788, MAE 748.0739), and Japan at 0.01 (RMSE 1457.708, MAE 511.2254), with a range of 123.077. The limited variation in Australia (1.9\%) contrasts with US-States (23.7\%), reflecting differing levels of sensitivity to this parameter.

\item \textbf{Alteration of RNN Dimension}
The RNN dimension, which affects the richness of hidden state representations, was tested at 10, 20, 30, 40, and 50. Findings, reported in Table \ref{tab:rnn_dimension}, indicate US-States reaching its lowest RMSE at 50 (233.0153, MAE 106.0453), Australia at 40 (RMSE 291.4734, MAE 90.3212), US-Regions at 30 (RMSE 1265.7188, MAE 722.9141), and Japan at 50 (RMSE 1601.9304, MAE 565.5046). Australia shows the widest RMSE range (538.7745), suggesting significant responsiveness, while US-Regions has the narrowest (165.7026). This implies that increasing dimensions can enhance performance, though excessive values may lead to overfitting in data-limited scenarios.
\end{itemize}
\textbf{Summary and Implications}:
The hyperparameter analysis highlights varied sensitivities across datasets and parameters. Key insights include:
\begin{itemize}
\item \textbf{Look-back Window}: US-States favors 50, Australia 20, and US-Regions and Japan 40. A window range of 20–50 is recommended to balance accuracy and efficiency.
\item \textbf{Filter Count}: Optimal at 16 (US-States), 8 (Australia), 4 (US-Regions), and 32 (Japan). A moderate count (4–32) is suggested to prevent overfitting.
\item \textbf{Learning Rate}: Best at 0.005 (US-States, US-Regions) and 0.01 (Australia, Japan), with a range of 0.005–0.01 proposed for effective training.
\item \textbf{RNN Dimension}: Peaks at 50 (US-States, Japan), 40 (Australia), and 30 (US-Regions), with 30–50 advised while monitoring overfitting risks.
\end{itemize}

\subsubsection{Interpretability Analysis}
\begin{figure}[H]
    \centering
    \begin{minipage}{0.45\textwidth}
        \includegraphics[width=\linewidth]{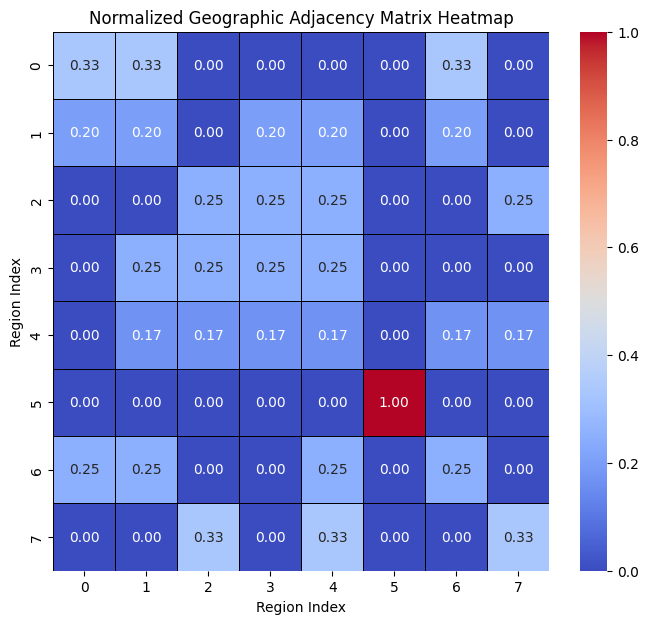}
        \caption{Normalized Geographic Adjacency Matrix Heatmap For Static Propagation}
        \label{fig: Normalized Geographic Adjacency Matrix Heatmap}
    \end{minipage}
    \hfill
    \begin{minipage}{0.45\textwidth}
        \includegraphics[width=\linewidth]{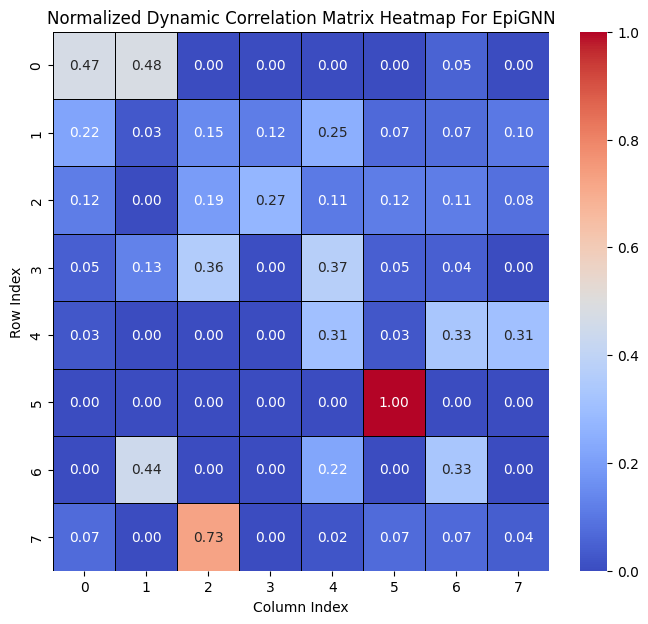}
        \caption{Normalized Dynamic Correlation Matrix Heatmap For EpiGNN}
        \label{fig: Normalized Dynamic Correlation Matrix Heatmap For EpiGNN}
    \end{minipage}

    \vspace{1cm} 

    \begin{minipage}{0.45\textwidth}
        \includegraphics[width=\linewidth]{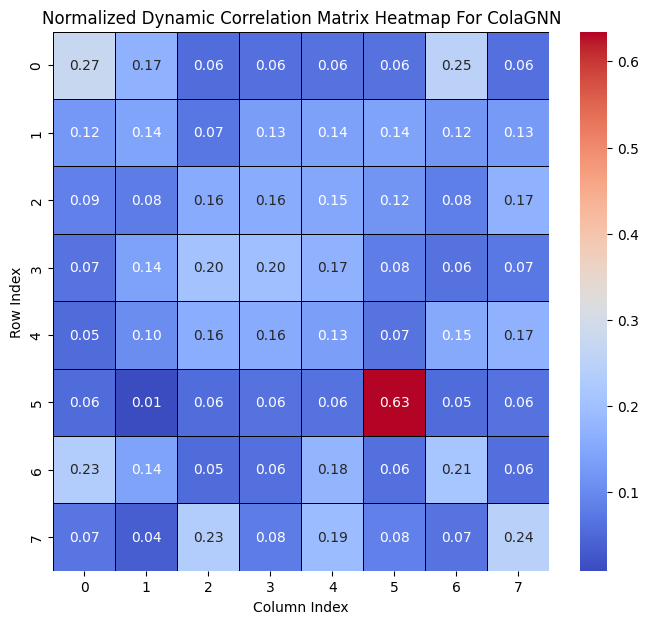}
        \caption{Normalized Dynamic Correlation Matrix Heatmap For ColaGNN}
        \label{fig: Normalized Dynamic Correlation Matrix Heatmap For ColaGNN}
    \end{minipage}
    \hfill
    \begin{minipage}{0.45\textwidth}
        \includegraphics[width=\linewidth]{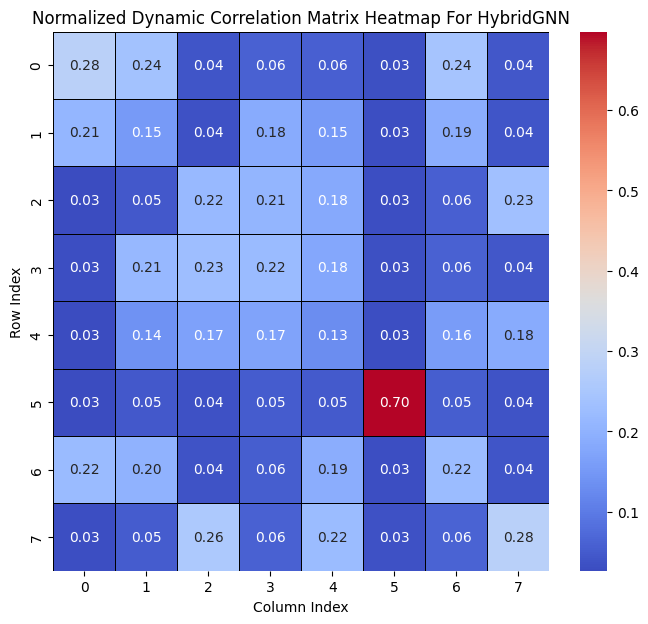}
        \caption{Normalized Dynamic Correlation Matrix Heatmap For HybridGNN}
        \label{fig: Normalized Dynamic Correlation Matrix Heatmap For HybridGNN}
        \end{minipage} 
\end{figure}

Based on the provided Figures, here is an analysis of why HybridGNN performs better and more stably than the other models on the Australian dataset (e.g., Horizon = 20). These heatmaps compare the performance of four different Graph Neural Network (GNN) models on an Australian dataset by analyzing their normalized correlation matrices. The four figures are:

\begin{itemize}
    
\item \textbf{Normalized Geographic Adjacency Matrix Heatmap} (see Figure~\ref{fig: Normalized Geographic Adjacency Matrix Heatmap}): This map shows the region adjacency matrix based on geographic locations. It is obtained by normalizing the matrix $\mathbf{A}^{geo}$ and reflects the physical proximity between regions.

\item \textbf{Normalized Dynamic Correlation Matrix Heatmap For EpiGNN} (see Figure~\ref{fig: Normalized Dynamic Correlation Matrix Heatmap For EpiGNN}): This heatmap displays the dynamic correlation matrix for the EpiGNN model, which attempts to capture time-varying interactions between regions.

\item \textbf{Normalized Dynamic Correlation Matrix Heatmap For ColaGNN} (see Figure~\ref{fig: Normalized Dynamic Correlation Matrix Heatmap For ColaGNN}): This heatmap shows the dynamic correlation matrix for the ColaGNN model.

\item \textbf{Normalized Dynamic Correlation Matrix Heatmap For HybridGNN} (see Figure~\ref{fig: Normalized Dynamic Correlation Matrix Heatmap For HybridGNN}): This heatmap shows the dynamic correlation matrix for the HybridGNN model.
\end{itemize}

To analyze why HybridGNN's predictions are better and more stable, we need to compare the characteristics of these four heatmaps, especially the differences between HybridGNN and the others. From the heatmaps provided, we can draw the following analysis:

\begin{enumerate}  

 \item\textbf{Normalized Geographic Adjacency Matrix}:

This map (see Figure~\ref{fig: Normalized Geographic Adjacency Matrix Heatmap}) shows region connectivity based on geographic location. For example, the element at row 5, column 5 is 1.00, indicating a high self-correlation for region 5 (which is expected). Other non-zero values indicate adjacent regions. This matrix is static; it only reflects spatial relationships and does not account for dynamic changes over time.

\item \textbf{ EpiGNN's Dynamic Correlation Matrix}:

This map (see Figure~\ref{fig: Normalized Dynamic Correlation Matrix Heatmap For EpiGNN}) displays the dynamic correlations learned by the EpiGNN model, which attempts to capture region connections in dynamic processes like disease spread. Compared to the geographic adjacency matrix, its pattern is different. For instance, the value at row 7, column 2 is 0.73, which is much higher than the corresponding value in the geographic adjacency matrix (0.00). This suggests that EpiGNN might have captured strong correlations between non-geographically adjacent regions. However, the matrix contains many zero values, which could mean the model focuses only on a few strongly correlated regions and ignores other potentially weaker correlations.

\item \textbf{ColaGNN's Dynamic Correlation Matrix}:

This map (see Figure~\ref{fig: Normalized Dynamic Correlation Matrix Heatmap For ColaGNN}) also shows dynamic correlations. Its values are generally smaller, except for row 5, column 5 (0.63). This might indicate that ColaGNN's ability to capture dynamic correlations is relatively weak, or that it distributes weights more evenly across correlations. The value distribution in this matrix appears more uniform than EpiGNN's, with fewer zero values. This could mean it considers more regional relationships but may not highlight the most important ones.

\item \textbf{HybridGNN's Dynamic Correlation Matrix}:

A key feature of this map (see Figure~\ref{fig: Normalized Dynamic Correlation Matrix Heatmap For HybridGNN}) is the high value at row 5, column 5 (0.70), similar to the self-correlation in the geographic adjacency matrix. Compared to the other models' dynamic correlation matrices, HybridGNN's value distribution seems smoother with more non-zero entries. The crucial point is that HybridGNN combines geographic adjacency information with dynamic correlation information. Its matrix pattern might be closer to the geographic adjacency matrix than EpiGNN's, while still incorporating features of dynamic change. For example, its values are generally higher than ColaGNN's, and it doesn't have an excessive number of zeros like EpiGNN.
\end{enumerate}

\textbf{Comprehensive Analysis and Conclusion}:

HybridGNN's superior and more stable performance on the Australian dataset is likely due to its hybrid approach, which effectively integrates static geographic adjacency information with dynamic temporal correlation information.
\begin{itemize}
   
 \item \textbf{Importance of Geographic Adjacency}: In prediction tasks like disease spread or traffic flow, geographically adjacent regions often have stronger mutual influences. Ignoring geographic location in favor of only dynamic correlations can cause a model to miss crucial foundational relationships.

\item \textbf{Importance of Dynamic Correlation}: Relying solely on a geographic adjacency matrix is static and fails to capture complex interactions that change over time (e.g., shifts in commuting patterns, changes in disease hotspots).

\item \textbf{HybridGNN's Advantage}: 
By combining these two types of information, HybridGNN can capture fundamental spatial dependencies, which are provided by the geographic adjacency matrix. It also can learn dynamic dependencies that change over time, which are provided by the temporal features in the data.
\end{itemize}
Therefore, the HybridGNN heatmap likely reflects a balance: it is neither a pure geographic adjacency matrix nor a completely random dynamic correlation matrix. It is probably a weighted combination or a clever fusion that better represents the true complex relationships between regions in the Australian dataset, leading to higher accuracy and stability in its predictions.

\section{Discussion of the EpiCola-GNN Model}
\subsection{Strengths of HybridGNN}
HybridGNN (EpiCola-GNN) distinguishes itself by merging and extending specific architectural components from both EpiGNN and ColaGNN. This design creates powerful synergies, offering capabilities and performance benefits beyond what its predecessor models can achieve individually.
\begin{itemize}
    \item \textbf{Complementary Dynamism:}
     While EpiGNN's GraphLearner generates a graph dynamic based on feature similarity, HybridGNN introduces the additional and distinct dynamism derived from temporal sequence modeling via an RNN. This allows HybridGNN to capture evolving relationships driven by explicit temporal causality (e.g., how the history of one region impacts its connection to another), which is different from a learned graph based solely on current feature interactions.
    \item \textbf{Richer Foundation for Propagation:}
     By integrating both the sequence-driven dynamic connections (from the ColaGNN side) and the feature-driven/geographical connections (from the EpiGNN side) into a single adjacency matrix, HybridGNN provides its GNN layers with an unparalleled, multi-faceted view of how regions truly interact and influence each other. This is critical for modeling complex propagation phenomena where both historical patterns and current states dictate connectivity.
    \item \textbf{Holistic Pre-GCN Feature Concatenation:}
    HybridGNN strategically prepares its initial node features for GCN propagation by concatenating (rather than summing) Multi-scale temporal features, Local transmission risk and Global transmission risk. This holistic concatenation, performed before GCN layers, provides the graph convolutional layers with an exceptionally rich, multi-faceted initial node representation. By explicitly preserving and processing the distinct contributions of temporal patterns, local risks, and global risks right from the start, the GCNs can learn more nuanced, contextually aware, and disentangled embeddings during message passing. This significantly enhances the GNN's ability to model complex spatio-temporal dependencies with greater precision.
    \item \textbf{Deeply Integrated Multi-Perspective Prediction Input:}
    HybridGNN's final prediction layer receives an input that uniquely combines its highly processed spatio-temporal graph output with a distinct, high-level temporal signal from an RNN.This final, deep fusion ensures the model's predictions are exceptionally well-informed, robust, and comprehensive. By leveraging both the deep, context-aware spatio-temporal insights from the GNN (representing the "where" and "what" of propagation) and the explicit, high-level temporal trajectory from the RNN (representing the "when" and "how fast"), HybridGNN can generate more accurate and reliable forecasts for dynamic systems, capturing both the fine-grained interactions and the overarching temporal trends.
\end{itemize}

\subsection{Future Improvements}
While HybridGNN's current architecture offers substantial advantages, certain refinements could further enhance its robustness and generalizability, particularly when facing highly volatile or complex system behaviors:
\begin{itemize}
    \item\textbf{Advanced Volatility Adaptation:}
    For scenarios characterized by extremely rapid and unpredictable shifts, the model's temporal components could benefit from dynamic weighting mechanisms that can instantly adjust the influence of short-term versus long-term features based on the current observed volatility. This would allow the model to automatically prioritize the most recent information during abrupt changes, preventing older, less relevant patterns from unduly influencing forecasts.
    \item\textbf{Deeper Inter-Module Feature Interaction:}
    The current approach to feature fusion, primarily concatenation, could be enhanced. Future work should explore more sophisticated architectural interfaces between modules. This includes designing cross-modal attention mechanisms that enable temporal features to directly modulate spatial aggregation in GCNs, or allowing risk encodings to dynamically refine the weights of connections in the dynamic graph construction process, leading to a more synergistic learning across components.
    \item\textbf{Structural Integration of Causal Inference:}
    To move beyond associative learning, a critical architectural evolution involves integrating explicit causal inference modules. This could entail designing specific layers that identify direct causal relationships between nodes or features, subsequently guiding the graph construction or message passing rules to prioritize these causal links. This would not only enhance prediction robustness but also provide invaluable, actionable insights into the underlying drivers of propagation.
     \item\textbf{Adaptive Component Contribution in Graph Fusion:}
     The dynamic graph is currently a summation of its parts. A more advanced approach would involve a learnable gating mechanism that dynamically weighs the contribution of each component (e.g., geographical, temporal, external relations) to the final graph structure. This would allow the model to autonomously emphasize the most relevant relational information based on the current context or prediction task, optimizing graph representation on the fly.
\end{itemize}

By systematically addressing these specific architectural refinements, HybridGNN can evolve into an even more versatile and reliable tool, offering increasingly precise and actionable forecasts that empower decision-makers in the face of diverse and unpredictable real-world challenges.

\section{Conclusion and Outlook}
EpiGNN and ColaGNN, as graph neural network-based regional epidemic forecasting models, have both significantly improved prediction accuracy and interpretability through innovative architectural designs. EpiGNN focuses on integrating transmission risk and external data, showing superiority in handling dynamic epidemics. ColaGNN, with its dynamic attention mechanism, excels in long-term forecasting and provides insights into propagation dynamics.

Building on this, this paper proposes HybridGNN (EpiCola-GNN), a novel model that combines EpiGNN's strengths in risk encoding and external data integration with ColaGNN's capabilities in long-term forecasting and dynamic attention mechanisms. Experimental results demonstrate that HybridGNN achieves superior prediction performance on multiple datasets (particularly Australia-COVID and US-States), outperforming individual models in short-term accuracy and overall stability, thus validating the effectiveness of our design philosophy in merging the advantages of both models.

Despite the significant progress made by HybridGNN, the field of epidemic forecasting still faces common challenges, such as generalization ability and model robustness on extremely volatile datasets. Future research can further enhance the model's robustness and practicality by introducing mechanisms for temporal decay, enhancing external data robustness, fusing multi-source data, and employing causal modeling. These advancements will provide more reliable support for public health decision-making. Continued research into these advanced GNN models, including our proposed HybridGNN, will keep driving the development of epidemic forecasting.

\end{document}